\documentclass[conf]{new-aiaa}
\usepackage[utf8]{inputenc}
\usepackage{svg}
\usepackage{graphicx}
\usepackage{amsmath}
\usepackage[version=4]{mhchem}
\usepackage{siunitx}
\usepackage{longtable,tabularx}
\usepackage{float}
\usepackage{caption}
\usepackage{subcaption}
\usepackage{booktabs} 
\usepackage{tablefootnote}
\usepackage[title]{appendix}
\hypersetup{colorlinks=true,linkcolor=blue,citecolor=blue}

\newcommand{\etal}{{\em et~al.}}
\usepackage{xcolor,pifont}
\newcommand*\colourcheck[1]{%
  \expandafter\newcommand\csname #1check\endcsname{\textcolor{#1}{\ding{52}}}%
}
\colourcheck{blue}
\colourcheck{green}
\colourcheck{red}
\newcommand*\colourxmark[1]{%
  \expandafter\newcommand\csname #1xmark\endcsname{\textcolor{#1}{\ding{55}}}%
}
\colourxmark{blue}
\colourxmark{green}
\colourxmark{red}

\title{OptiWing3D: A Diverse Dataset of Optimized Wing Designs}

\author{Cashen Diniz\footnote{Research Assistant and Ph.D. Student, Department of Mechanical and Process Engineering.} and Mark D. Fuge \footnote{Professor, Department of Mechanical and Process Engineering.}}
\affil{Department of Mechanical and Process Engineering, ETH Zürich, Rämistrasse 101, 8092 Zürich, Switzerland.}

\begin{document}

\maketitle

\begin{abstract}
    
OptiWing3D is the first publicly available dataset of high-fidelity shape optimized 3D wing geometries. Existing aerodynamics datasets are either limited to 2D simulations, lack optimization, or derive diversity solely from perturbations to a single baseline design, constraining their application as benchmarks to inverse design approaches and in the study of design diversity. The OptiWing3D dataset addresses these gaps, consisting of 1552 simulations resulting in 776 wing designs initialized from distinct extruded airfoil cross-sections. Additionally, a majority of the optimized wings in the dataset are paired to 2D counterparts optimized under identical conditions, creating the first multi-fidelity aerodynamic shape optimization dataset. Moreover, this structure allows for a direct comparison between 2D and 3D aerodynamic simulations. It is observed that 3D optimized designs diverge most prominently from the 2D-optimized designs near the wingtip, where three-dimensional effects are strongest, a finding made possible by the paired nature of the dataset. Finally, we demonstrate a constraint-aware conditional latent diffusion model capable of generating optimized wings from flow conditions, establishing a baseline for future inverse design approaches. The dataset, containing wing geometries and surface pressure distributions is publicly released to advance research in data-driven aerodynamic design \footnote{https://github.com/cashend/OptiWing3D}. 
    
\end{abstract}

\section{Introduction and Related Work}

In aerodynamic design, intuition about the optimal shape of a lifting body given a set of flow conditions and constraints could drastically decrease the cost of expensive physical experimentation, and reduce reliance on high-fidelity computational fluid dynamics (CFD) simulations. Despite this, such an intuition is typically formed through experience and may be difficult to transfer between individuals or encode in design tools. 

Although it is difficult to completely replace learned intuition, one approach which could help alleviate this gap is to leverage existing simulation or experimental data to develop more accessible surrogate models or generative methods that are able to better guide early-stage design decisions. This so-called 'data-driven' approach has gained traction in the field of aerodynamic design. 

Li~\etal~presented a data-based method for using RANS simulations to train a machine learning-based model to predict the lift and drag for two different flow conditions (transonic and subsonic) for a variety of airfoil shapes. Li~\etal~demonstrated that such surrogate models could be used to partially replace expensive adjoint-based optimizers for certain 2D aerodynamic shape optimization problems~\cite{LiData, li2022machine}. In addition, other data-based approaches have taken advantage of advances in generative machine learning. For example, Chen~\etal~ presented an optimized dataset of airfoil designs that could be used to train a generative adversarial network (GAN) to produce optimized airfoil designs. Despite presenting valuable data on optimized designs as a function of a broad range of flow conditions, the authors still conceded issues with the dataset design diversity and constraints ~\cite{chen2021inverse}. Other authors have also used GANs for the parameterization of airfoil geometries, and the prediction of the flow fields around them~\cite{wang2023airfoil, haizhou2022generative}. More recently, other generative models have found use, including invertible neural networks (INNs), used to inversely estimate airfoil shapes given performance, and diffusion models, used to produce optimal airfoil designs~\cite{glaws2022invertible, diniz2024optimizing}. Despite the immense promise of fast early-stage airfoil design that these models may provide, they are inherently limited by the data they are trained on, in terms of optimality, data size, fidelity, and diversity. Unfortunately, publicly available aerodynamic design datasets that satisfy all or even some of these criteria are scarce or non-existent.

Table~\ref{tab:datasets} describes a selection of some of the more prominent existing aerodynamic design datasets. We highlight attributes relevant to data-driven aerodynamic shape design. We organize our discussion around three attributes: geometric diversity, optimization, and dimensionality.

\begin{table}[H]
\caption{Related Aerodynamic Datasets}
\label{tab:datasets}
\resizebox{\textwidth}{!}{%
\begin{tabular}{@{}lllllllll@{}}
\toprule
\multicolumn{1}{c}{Dataset} &
  \multicolumn{1}{c}{\begin{tabular}[c]{@{}c@{}}Distinct Initial\\ Designs\end{tabular}} &
  Size &
  \multicolumn{1}{c}{Optimized} &
  \multicolumn{1}{c}{\begin{tabular}[c]{@{}c@{}}Primary \\ Design Type\end{tabular}} &
  \multicolumn{1}{c}{\begin{tabular}[c]{@{}c@{}}Design \\ Dimensionality\end{tabular}} &
  \multicolumn{1}{c}{\begin{tabular}[c]{@{}c@{}}Diverse Shape \\ Initialization\end{tabular}} &
  \multicolumn{1}{c}{\begin{tabular}[c]{@{}c@{}}Sampled \\ Flow Conditions\end{tabular}} &
  \multicolumn{1}{c}{Solver} \\ \midrule
\begin{tabular}[c]{@{}l@{}}PALMO\\ \cite{cornelius2025palmo}\end{tabular} &
  16 NACA airfoils &
  \begin{tabular}[c]{@{}l@{}}53k \\ simulations\end{tabular} &
  \redxmark &
  Airfoils &
  2D &
  \redxmark &
  \begin{tabular}[c]{@{}l@{}}M $\sim${[}0.25, 0.90{]}\\ Re $\sim$8 distinct,\\ {[}75E3, 8E6{]}\end{tabular} &
  \begin{tabular}[c]{@{}l@{}}Compressible \\ steady RANS\\ (OVERFLOW)\end{tabular} \\
\begin{tabular}[c]{@{}l@{}}Airfoil CFD - 2k \\ \cite{OEDI_Dataset_5970}\end{tabular} &
  1800 airfoils &
  \begin{tabular}[c]{@{}l@{}}250k\\ simulations\end{tabular} &
  \redxmark &
  Airfoils &
  2D &
  \greencheck &
  \begin{tabular}[c]{@{}l@{}}Single M = 0.1,\\ Re $\sim$3 distinct,\\ {[}3E6, 9E6{]}\end{tabular} &
  \begin{tabular}[c]{@{}l@{}}Compressible\\ steady RANS\\ (HAM2D)\end{tabular} \\
\begin{tabular}[c]{@{}l@{}}Chen~\etal\\ \cite{chen2021inverse}\end{tabular} &
  \begin{tabular}[c]{@{}l@{}}8 distinct initial \\ airfoils\end{tabular} &
  \begin{tabular}[c]{@{}l@{}}1245 \\ optimized airfoils \end{tabular} &
  \greencheck &
  Airfoils &
  2D &
  \redxmark &
  \begin{tabular}[c]{@{}l@{}}M $\sim${[}0.2, 0.90{]}\\ Re $\sim${[}1E7, 1E8{]}\end{tabular} &
  \begin{tabular}[c]{@{}l@{}}Compressible\\ steady RANS\\ (SU2)\end{tabular} \\
\begin{tabular}[c]{@{}l@{}}DrivAerNet++\\ \cite{elrefaie2024drivaernet++}\end{tabular} &
  \begin{tabular}[c]{@{}l@{}}8000 distinct car \\ configurations\end{tabular} &
  \begin{tabular}[c]{@{}l@{}}8000\\ simulations\end{tabular} &
  \redxmark &
  Cars &
  3D &
  \greencheck &
  \begin{tabular}[c]{@{}l@{}}Single M = 0.09\\ Re $\sim${[}8.36E6, 1E7{]}\end{tabular} &
  \begin{tabular}[c]{@{}l@{}}Incompressible \\ steady RANS \\ (OpenFOAM)\end{tabular} \\
AWSD \cite{su2023awsd} &
  \begin{tabular}[c]{@{}l@{}}300 distinct wing\\ parameterizations, \\ single AoA\end{tabular} &
  \begin{tabular}[c]{@{}l@{}}300\\ simulations\end{tabular} &
  \redxmark &
  Wings &
  3D &
  \redxmark &
  \begin{tabular}[c]{@{}l@{}}Single M = 0.95\\ (Euler equations)\end{tabular} &
  \begin{tabular}[c]{@{}l@{}}Compressible \\ steady Euler \\ equations\end{tabular} \\
\begin{tabular}[c]{@{}l@{}}Diniz~\etal\\ \cite{diniz2024optimizing}\end{tabular} &
  \begin{tabular}[c]{@{}l@{}}935 distinct initial\\ airfoils\end{tabular} &
  \begin{tabular}[c]{@{}l@{}}935 \\ optimized airfoils,\\ (~75k simulations)\end{tabular} &
  \greencheck &
  Airfoils &
  2D &
  \greencheck &
  \begin{tabular}[c]{@{}l@{}}M $\sim${[}0.4, 0.9{]}\\ Re $\sim${[}1E6, 1E7{]}\end{tabular} &
  \begin{tabular}[c]{@{}l@{}}Compressible \\ steady RANS  \\ (ADFlow)\end{tabular} \\
\begin{tabular}[c]{@{}l@{}}BlendedNet\\ \cite{sung2025blendednet}\end{tabular} &
  \begin{tabular}[c]{@{}l@{}}1 base BWB geometry, \\ 999 distinct variations (no local\\ shape variation) \end{tabular} &
  \begin{tabular}[c]{@{}l@{}}8830\\ simulations\end{tabular} &
  \redxmark &
  \begin{tabular}[c]{@{}l@{}}Blended\\ Wing Body\end{tabular} &
  3D &
  \redxmark &
  \begin{tabular}[c]{@{}l@{}}M $\sim${[}0.05, 0.5{]}\\ L (Re) $\sim${[}0.1m, 10m{]}\end{tabular} &
  \begin{tabular}[c]{@{}l@{}}Compressible\\ steady RANS\\ (FUN3D)\end{tabular} \\
\begin{tabular}[c]{@{}l@{}}Emmi-Wing\\ \cite{paischer2025emmi}\end{tabular} &
  \begin{tabular}[c]{@{}l@{}}1 base NACA0012 extruded\\ wing, 30k variations (no local\\ shape variation)\end{tabular} &
  \begin{tabular}[c]{@{}l@{}}30k \\ simulations\end{tabular} &
  \redxmark &
  Wings &
  3D &
  \redxmark &
  \begin{tabular}[c]{@{}l@{}}U$_\infty$ $\sim${[}150, 300{]} m/s \\ Single Re (implied) \\ \end{tabular} &
  \begin{tabular}[c]{@{}l@{}}Compressible\\ steady RANS\\ (OpenFOAM)\end{tabular} \\
  \bf{OptiWing3D} &
  \begin{tabular}[c]{@{}l@{}}776 distinct initial\\ wings\end{tabular} &
  \begin{tabular}[c]{@{}l@{}}776 optimized wings, \\(1552 simulations\tablefootnote{From initial design simulations - or in the case where the first evaluation failed - the first successful evaluation. An additional $\sim$40k intermediate evaluations have  not been included in the current dataset version.}) \\ \end{tabular} &
  \greencheck &
  Wings &
  3D &
  \greencheck &
  \begin{tabular}[c]{@{}l@{}}M $\sim${[}0.4, 0.9{]}\\ Re $\sim${[}1E6, 1E7{]}\end{tabular} &
  \begin{tabular}[c]{@{}l@{}}Compressible \\ steady RANS \\ (ADFlow)\end{tabular} \\ \bottomrule
\end{tabular}%
}
\end{table}

\textbf{Simulation-only datasets.} Several datasets provide aerodynamic simulations across varied geometries or flow conditions but do not include optimized designs. PALMO, a closed-source benchmarking dataset, offers extensive flow condition coverage across 16 NACA airfoils, while Airfoil CFD-2k spans approximately 1800 distinct airfoil shapes but at a single Mach number and 3 distincr Reynolds numbers~\cite{cornelius2025palmo, OEDI_Dataset_5970}. In 3D, DrivAerNet++ provides diverse automotive geometries, and AWSD and Emmi-Wing offer wing simulations. Even so, AWSD is limited to inviscid Euler equations at a single, zero angle-of-attack. Emmi-Wing does provide variations for global design parameters, such as angle-of-attack, span, taper, and sweep, but there is no local shape variation, as all variations derive from a single extruded NACA0012 airfoil. Additionally, the simulations are conduced at a single (implied) Reynolds number~\cite{elrefaie2024drivaernet++, su2023awsd, paischer2025emmi}. BlendedNet similarly generates variations from a single blended-wing-body baseline~\cite{sung2025blendednet}. While valuable for surrogate modeling, these datasets are unable to serve as ground-truth benchmarks for inverse design methods, which require known optimal geometries for given performance targets and constraints.

\textbf{Optimized datasets.} Datasets containing optimized designs enable direct benchmarking of generative and inverse design approaches. Chen~\etal~presented optimized airfoils across a broad range of flow conditions; however, the authors noted that many optimized shapes remained similar to their initializations, and only 8 distinct starting geometries were used, which given optimizer variable bounds, limits design diversity~\cite{chen2021inverse}. Our previous work addressed diversity by initializing from 935 distinct airfoils, each optimized under unique flow conditions with explicit area and thickness constraints~\cite{diniz2024optimizing}. Both datasets, however, are restricted to 2D.

\textbf{A Gap: 3D Optimized Datasets.} To our knowledge, no publicly available dataset contains high-fidelity optimized 3D aerodynamic geometries. The gap between 2D and 3D simulations is prominent. While 2D simulations can offer computational efficiency, they also neglect three-dimensional effects such as tip vortices, induced drag, and spanwise pressure gradients. All of these effects influence realistic wing performance, and thus the optimal designs. Existing 3D datasets lack optimization entirely, and derive geometric variation solely from global design variables rather than local shape variation. 

OptiWing3D addresses these shortcomings, being the first publicly available dataset of high-fidelity, shape-optimized 3D wings, initialized from 776 distinct airfoil cross-sectional extrusions. Additionally, a majority of the designs are paired with 2D counterparts optimized under identical conditions, enabling multi-fidelity studies and a direct comparison between 2D and 3D optimization outcomes.

The key contributions of this paper are:
\begin{enumerate}
    \item OptiWing3D, the first publicly available dataset of high-fidelity, shape-optimized 3D wing geometries. The dataset has been paired with a 2D optimized airfoil dataset, allowing for future multi-fidelity studies. This structure lays the groundwork for; 1) direct comparison of 2D and 3D aerodynamic shape optimization and simulations, 2) benchmarking of inverse design methods on gradient-optimized ground truth, and 3) multi-fidelity transfer learning studies. 
    \item A demonstration and benchmarking of a diffusion generative model that is able to produce accurate optimized designs, given a set of flow conditions and constraints. We build this model to aid in future comparisons to other generative models built on the dataset presented in this work. 
\end{enumerate}

\section{Dataset Generation}

\subsection{Dataset Summary and Structure}

The OptiWing3D is composed of shape-optimized 3D wings, derived from a diverse initialization of extruded 2D airfoils. Additionally, OptiWing3D is paired to a 2D dataset collected as part of previous work~\cite{diniz2024optimizing}, composed of optimized 2D airfoil designs, initialized from the same airfoil geometries. In this case, each optimization is conducted under the same flow conditions and constraints, although it should be noted that because of optimizer and simulation failures, only a subset of the two datasets match. 

Although more data may be made available in the future, the current OptiWing3D dataset is structured to contain: 

\begin{enumerate}
    \item Slices along the span of both the optimized and initial geometries.
    \item Corresponding surface coefficient of pressure, and velocity distributions.
    \item Aerodynamic coefficients including coefficient of lift, $C_{L}$ and drag, $C_{D}$. 
    \item Input flow conditions, such as Mach, $M$, Reynolds number, $Re$, and constraints, such as the required lift coefficient, $C_{L}^{con}$ and minimum volume fraction ratio , $\left ( \frac{V}{V_{init}} \right )_{min}$.
\end{enumerate}

It should also be mentioned that the optimizer was permitted to introduce slight dihedral variation, $\eta_{y}$, by leaving the trailing-edge unconstrained along the spanwise (y) axis. This feature is embedded in the geometry slices, but is noted here for additional clarity.

\subsection{Design Conditions}

We define a diverse optimization problem specified by the flow conditions; Mach number ($M$) and Reynolds number ($Re$). We impose an equality  constraint specifying a target coefficient of lift ($C_{L}$ ) while keeping the volume above the minimum volume $\left ( \left( \frac{V}{V_{init}} \right)_{min} \right)$ constraint. Latin-hypercube sampling was used to generate the input parameter space, within the bounds specified in Table~\ref{tab:sampling}. Note that these are the same conditions specified for the paired 2D dataset~\cite{diniz2024optimizing}.

\begin{table}[H]
\caption{Sampled Parameter Bounds}
\centering
\label{tab:sampling}
\begin{tabular}{lllll}
\hline
\textbf{Category}       & \textbf{Parameter}                        & \textbf{Lower} & \textbf{Upper} & \textbf{Description}  \\ \hline
\textbf{Flow Condition} & M                                         & 0.4            & 0.9            & Mach Number           \\
                        & Re                                        & 1E6            & 10E6           & Reynolds Number      \\ \hline
\textbf{Constraint}     & $C_{L}^{con}$                             & 0.5            & 1.2            & Coefficient of Lift   \\
                        & $\left( \frac{V}{V_{init}} \right)_{min}$ & 0.75           & 1.0            & Minimum Volume Fraction \\ \hline
\end{tabular}
\end{table}

Additionally, we run all simulations at $T=300K$, and all optimizations were initialized using an initial angle of attack, $\alpha_{0} = 2.5^{\circ}$. 

\subsection{Wing Parameterization}

\begin{figure}[H]
    \centering
    \includegraphics[scale=0.30]{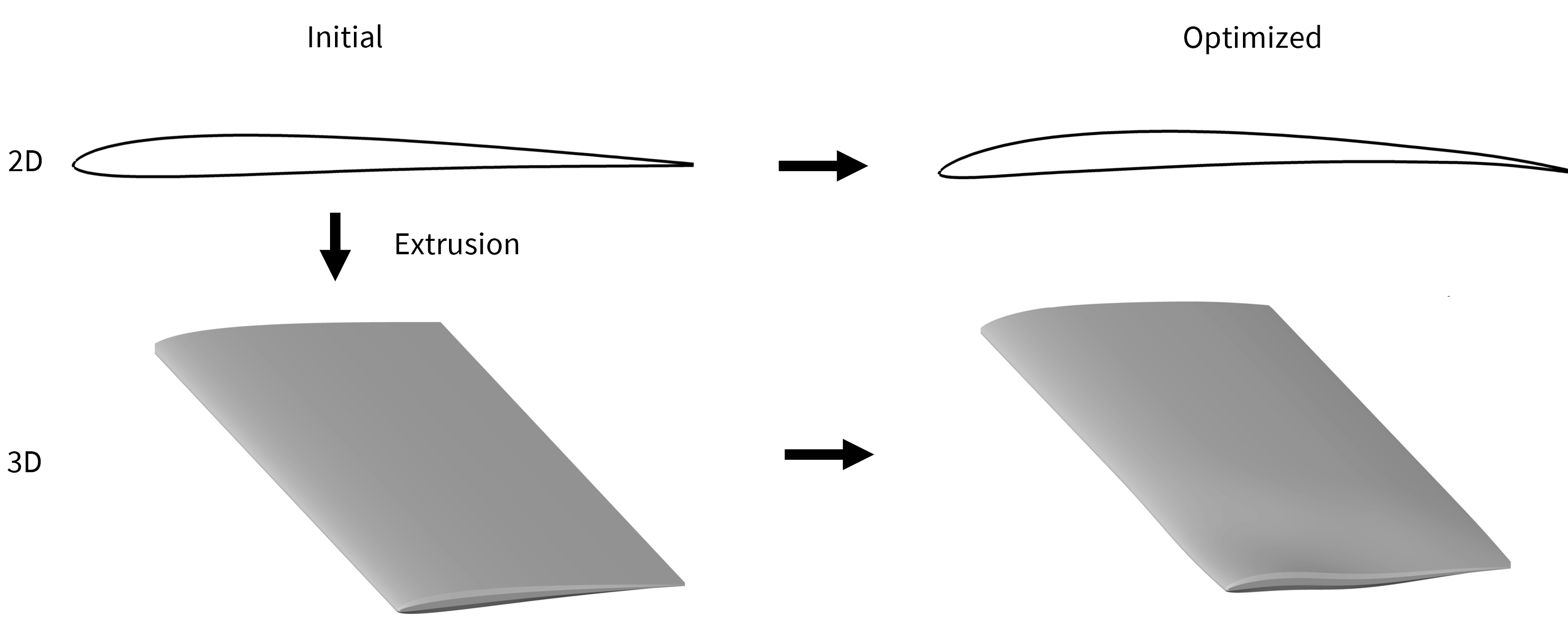}
    \caption{2D Airfoil and 3D Wing Data Generation}
    \label{fig:generation}
\end{figure}

We parameterized all wings as extrusions of 1400 initial airfoils taken from a random subset of the UIUC airfoil dataset~\cite{UIUC}. Each airfoil had a chord length of 1m, and was extruded a distance of 2.5m. Although the physical dimensions of the wings are fixed, the dataset remains generalizable to larger geometries through Reynolds number similarity.

\begin{figure}[H]\centering
\includegraphics[width=0.50\textwidth]{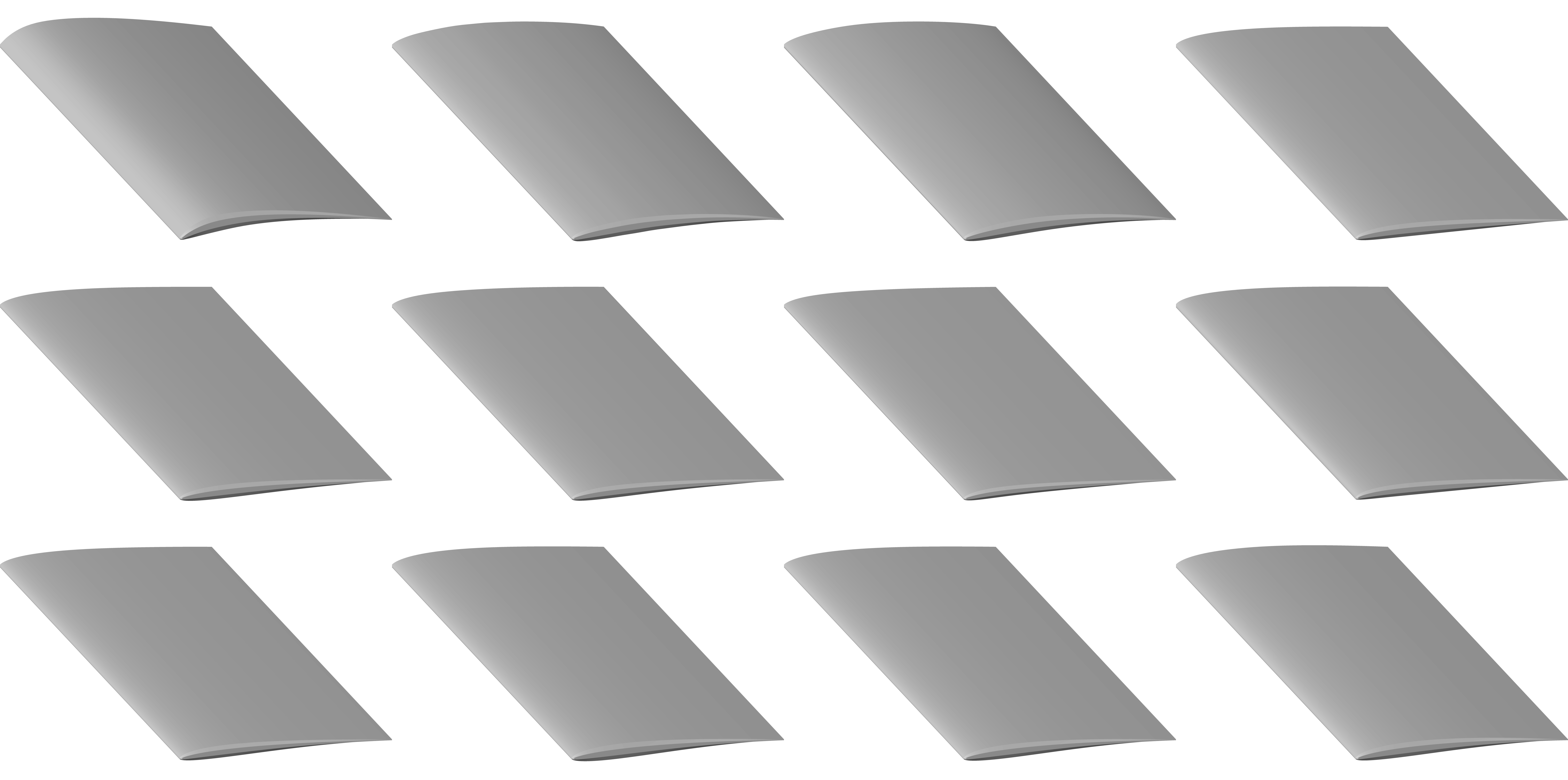}
\caption{Selected Initial (Non-Optimized) Wing Designs From the OptiWing3D Dataset}
\label{fig:nonopt}
\end{figure}

The raw coordinates in the dataset were resampled and smoothed. A simplified diagram of the process is shown in Figure~\ref{fig:generation}. A random set of initial extruded wings is presented in Figure~\ref{fig:nonopt}. We note that due to CFD convergence issues, infeasibility in optimizations, and failed meshes, only 776 of the 1400 cases were successfully optimized. Here we counted successful optimizations as satisfying a convergence accuracy of 1E-6.

\subsection{Mesh Generation and Deformation}

As each optimization starts from a unique extruded geometry, rather than a single baseline geometry (as is done in related work), a new mesh had to be generated for every optimization. Since ADFlow uses a structured meshing approach, it was not possible to use any of the open-source automated mesh generators, as these are, to our knowledge, restricted to unstructured meshes. To overcome this, we implemented a custom structured surface mesh generation pipeline for wings in gmsh, an open-source suite of mesh generating tools~\cite{gmsh}. 

\begin{figure}[H]
    \centering
    \begin{subfigure}[b]{0.495\textwidth}
        \centering
        \includegraphics[width=\textwidth]{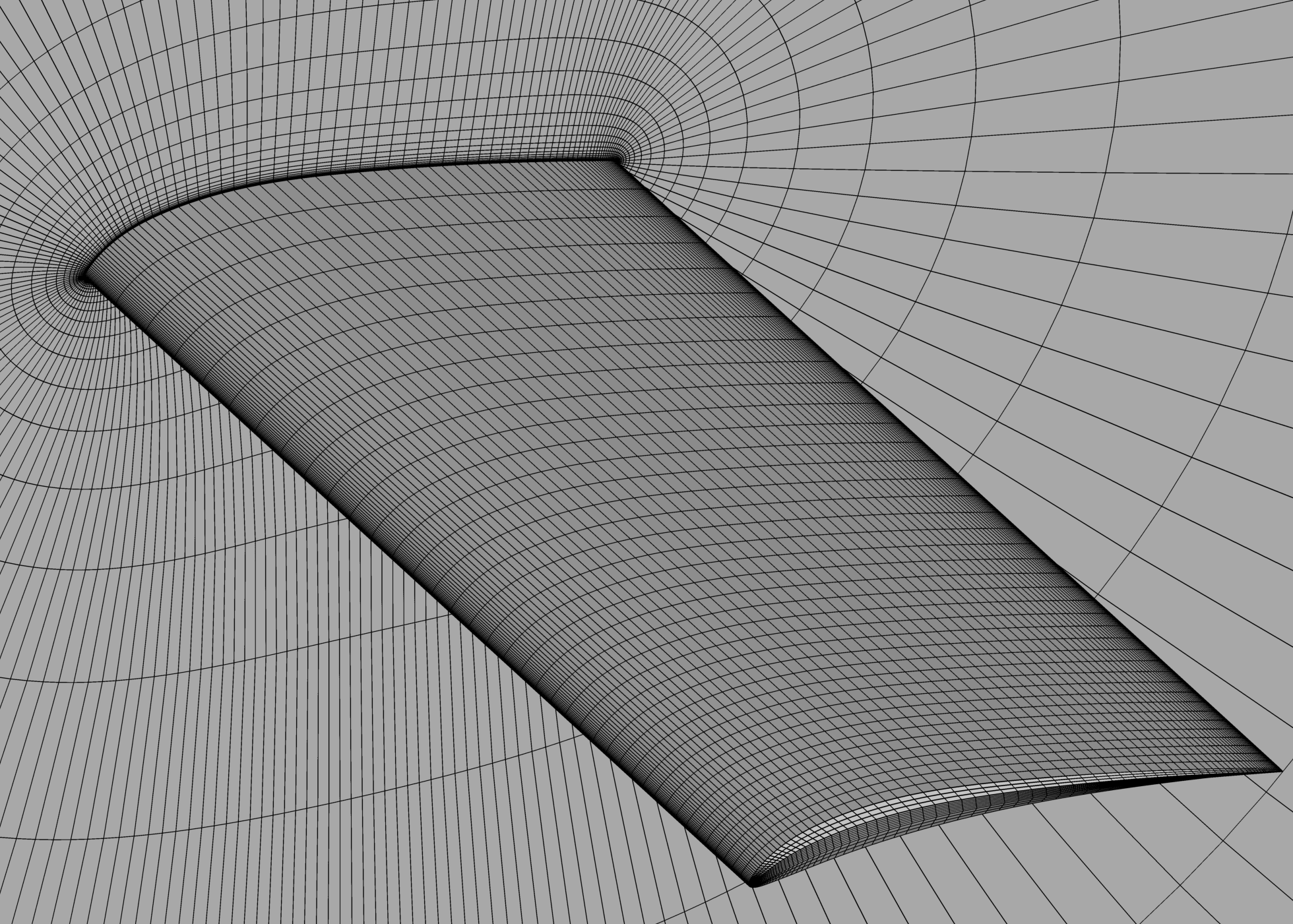}
        \caption{Sample Generated Surface Mesh}
        \label{fig:mesh}
    \end{subfigure}
    \hfill
    \begin{subfigure}[b]{0.495\textwidth}
        \centering
        \includegraphics[width=\textwidth]{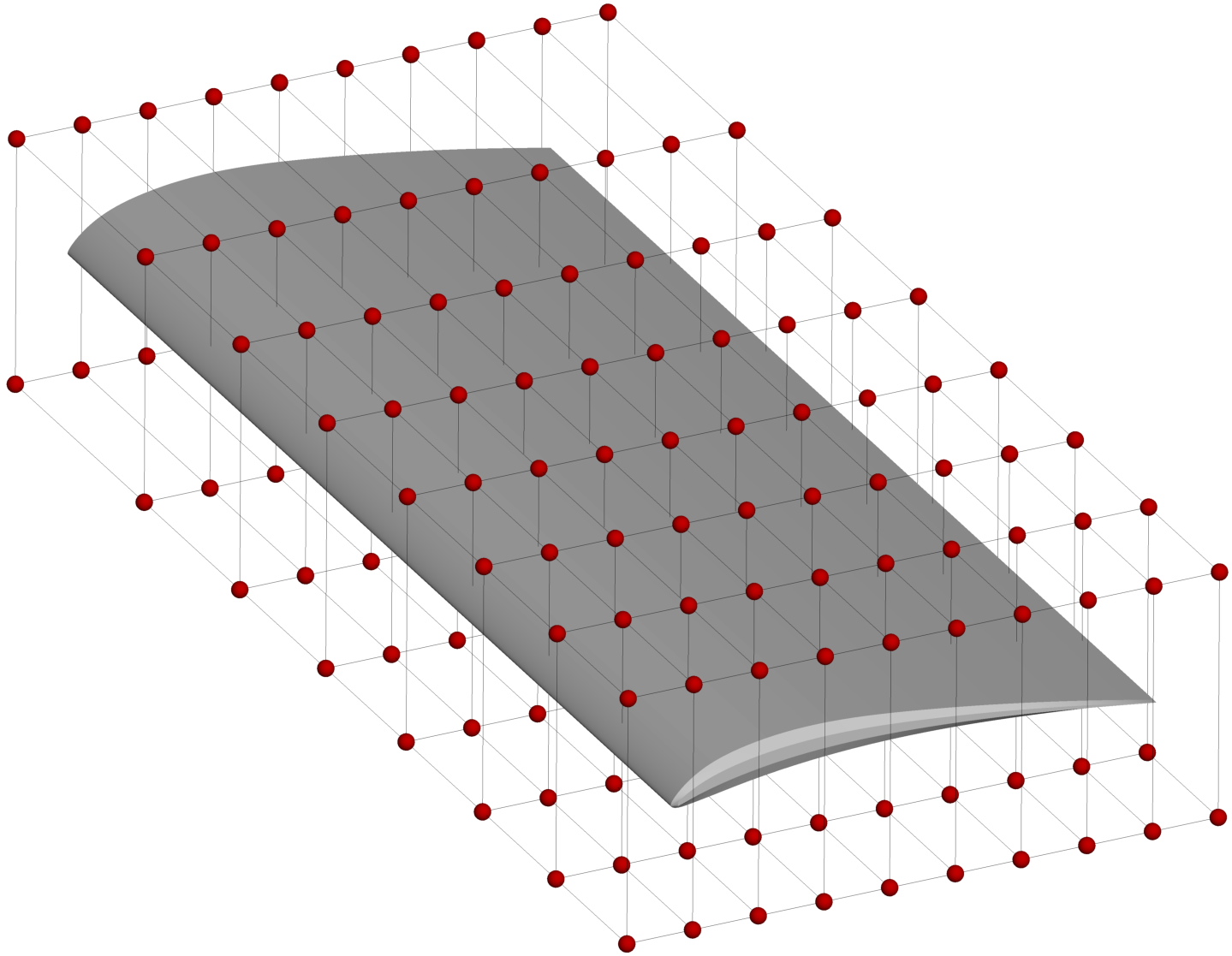}
        \caption{Sample FFD Cage Initialization}
        \label{fig:ffd_cage}
    \end{subfigure}
        \caption{Sample Mesh and FFD Cage}
        \label{fig:sample_mesh_ffd}
\end{figure}

Our algorithm uses the transfinite surface meshing capabilities in gmsh to extrude an airfoil cross-section into a wing. At the end of the wing, we employ a simple cut-wingtip treatment. The volume mesh extending into the freestream was generated using the hyperbolic mesh generator, pyHyp~\cite{PyHyp}. The total number of mesh cells used in each case is approximately 450k. Figure~\ref{fig:mesh} shows a sample surface mesh from the dataset. In addition, we use the ADFlow integrated Spalart-Allmaras turbulence model, and attempt to achieve a $y+$ value close to 1, estimated using the flow conditions and a flat-plate assumption (details can be found in Appendix A).

After the volume mesh is generated, changes to each initial mesh are parameterized by a free-form deformation (FFD) cage. The FFD cage allows for smooth and expressive deformations of the geometries in the optimization process. A rectangular prism of FFD points were constructed around the initial design, with allowable deformations restricted to the y-axis. Figure~\ref{fig:ffd_cage} shows the rectangular FFD cage initialized around a sample geometry from the dataset. In total, 160 FFD variables were used, $N_x = 10$ along the x-axis, $N_y = 2$ along the y-axis, and $N_{\eta} = 8$ along the spanwise axis.

\subsection{Optimization Problem}

We define the following optimization problem in Equation~\ref{eq:OptProblem}.

\begin{equation}
\begin{aligned}
\min_{\Delta y_{i},\alpha} \quad & C_{D}\\
\textrm{s.t.} \quad & C_{L} = C_{L}^{con}  \\
    & -0.025 \leq \Delta y_{i} \leq 0.025 \\
    & 0.0 \leq \alpha \leq 10.0 \\
    &\left( \frac{V}{V_{init}} \right)_{min} \leq \frac{V}{V_{init}} \leq 1.2 \\
    &0.15 \leq \frac{t}{t_{init}} \leq 3.0 \\
    &y_{TE}^{upper} = -y_{TE}^{lower} \\
    &y_{LE}^{upper} = -y_{LE}^{lower} \\
    \label{eq:OptProblem}
\end{aligned}
\end{equation}

Equation \eqref{eq:OptProblem} is purposefully similar to the optimization problem constructed for the paired 2D problem (see~\cite{diniz2024optimizing} for details), as to allow for as close to a one-to-one comparison as possible. A full list of the optimization variables is described in Table~\ref{tab:OptProblem}.

\begin{table}[H]
\centering
\caption{Optimization Problem Parameters}
\label{tab:OptProblem}
\resizebox{\columnwidth}{!}{%
\begin{tabular}{lllllll}
\hline
\textbf{Category}   & \textbf{Parameter}                 & \textbf{Quantity} & \textbf{Lower}                            & \textbf{Upper} & \textbf{Units}  & \textbf{Description}                                                                                              \\ \hline
\textbf{Objective}  & $C_{D}$                            & 1                 & -                                         & -              & Non-Dim./Counts & Coefficient of Drag                                                                                               \\ \hline
\textbf{Variable}   & $\Delta y_{i}$                     & 160                & -0.025                                    & 0.025          & m               & \begin{tabular}[c]{@{}l@{}} FFD point deformation in the y-axis direction, relative to initial value:\\ $\Delta y_{i} = y_{i} - y_{\textit{init}}$\end{tabular} \\
\textbf{}           & $\alpha$                           & 1                 & 0.0                                       & 10.0           & Degrees         & Angle of Attack                                                                                                   \\ \hline
\textbf{Constraint} & $C_{L} = C_{L}^{con}$              & 1                 & 0.0                                       & 0.0            & Non-Dim.        & Coefficient of Lift                                                                                               \\
                    & $\frac{V}{V_{init}}$               & 1                 & $\left( \frac{V}{V_{init}} \right)_{min}$ & 1.20           & Non-Dim.        & Volume Fraction; Relative to Initial.                                                                               \\
                    & $\frac{t}{t_{init}}$               & 200               & 0.15                                      & 3.00           & Non-Dim.        & Thickness Fraction; Relative to Initial                                                                           \\
                    & $y_{TE}^{upper} = -y_{TE}^{lower}$ & 8                 & 0.0                                       & 0.0            & m               & \begin{tabular}[c]{@{}l@{}}Trailing Edge FFD Point Shearing \\ Twist Condition\end{tabular}                       \\
                    & $y_{LE}^{upper} = -y_{LE}^{lower}$ & 8                 & 0.0                                       & 0.0            & m               & \begin{tabular}[c]{@{}l@{}}Leading Edge FFD Point Shearing\\ Twist Condition\end{tabular}                        
\end{tabular}%
}
\end{table}

\subsection{Solver}

All simulations were performed inside the MACH-Aero aerodynamic optimization framework.\footnote{\url{https://github.com/mdolab/MACH-Aero}.} ADflow, a differentiable CFD solver, and a component of the MACH-Aero framework was used to run the RANS simulations themselves ~\cite{ADFLOW, ADFLOWADJOINT}. ADFlow was set up to use the Spalart-Allmaras turbulence model as well as the approximate Newton-Krylov (ANK) method for increased robustness~\cite{ANK}. The sequential least squares programming algorithm (SLSQP) was used inside the pyOptSparse optimization framework~\cite{PyOptSparse}. Finally, the initial mesh deformation and parameterization of the FFD points were completed using the pyGeo and IDWarp frameworks~\cite{PyGeo, PyHyp}.

\section{Dataset Results}

\subsection{Optimized Designs}

\begin{figure}[H]
    \centering
    \begin{subfigure}[b]{0.495\textwidth}
        \centering
        \includeinkscape[width=\textwidth]{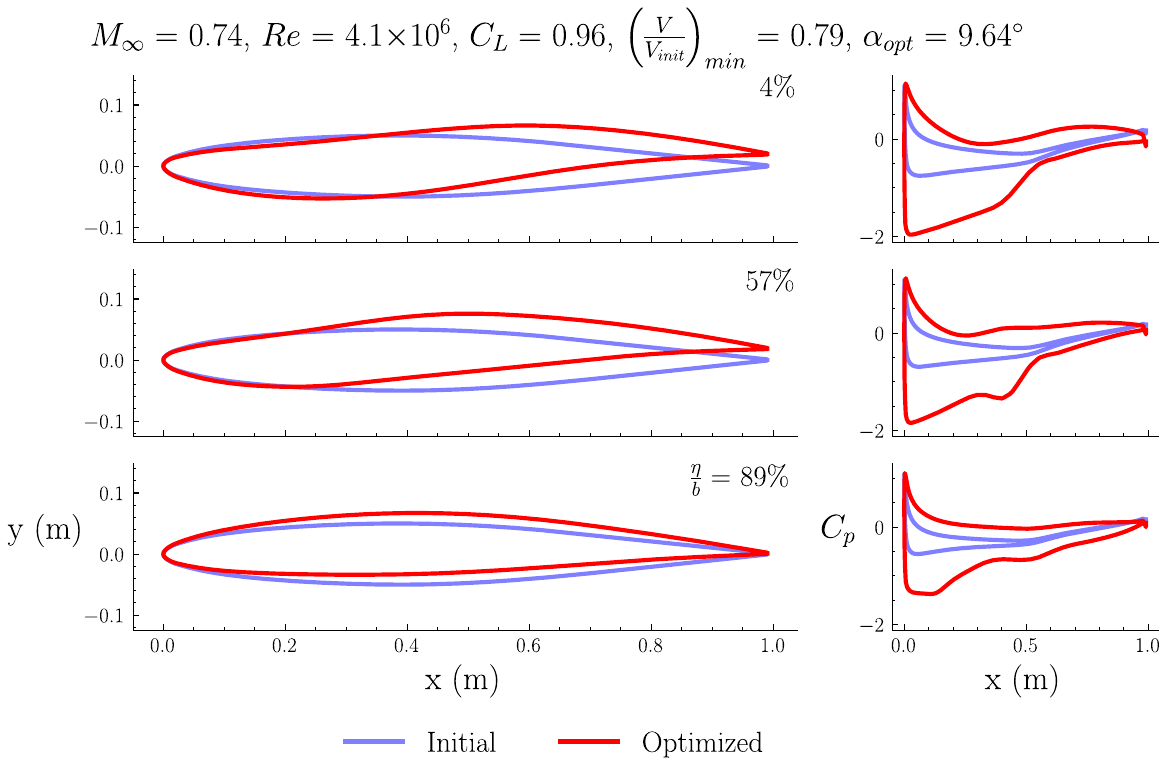}
        \caption{Sample a}
        \label{fig:sample_a_opt_init}
    \end{subfigure}
    \hfill
    \begin{subfigure}[b]{0.495\textwidth}
        \centering
        \includeinkscape[width=\textwidth]{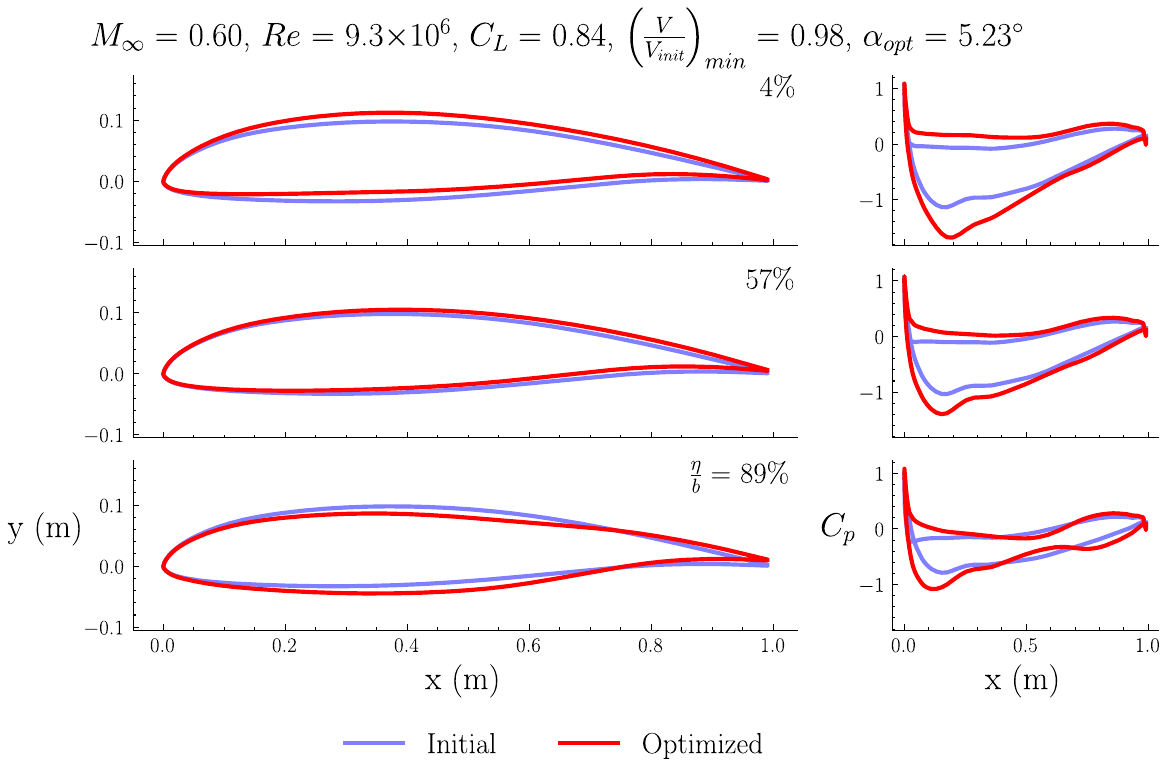}
        \caption{Sample b}
        \label{fig:sample_b_opt_init}
    \end{subfigure}
    \hfill
    \begin{subfigure}[b]{0.495\textwidth}
        \centering
        \includeinkscape[width=\textwidth]{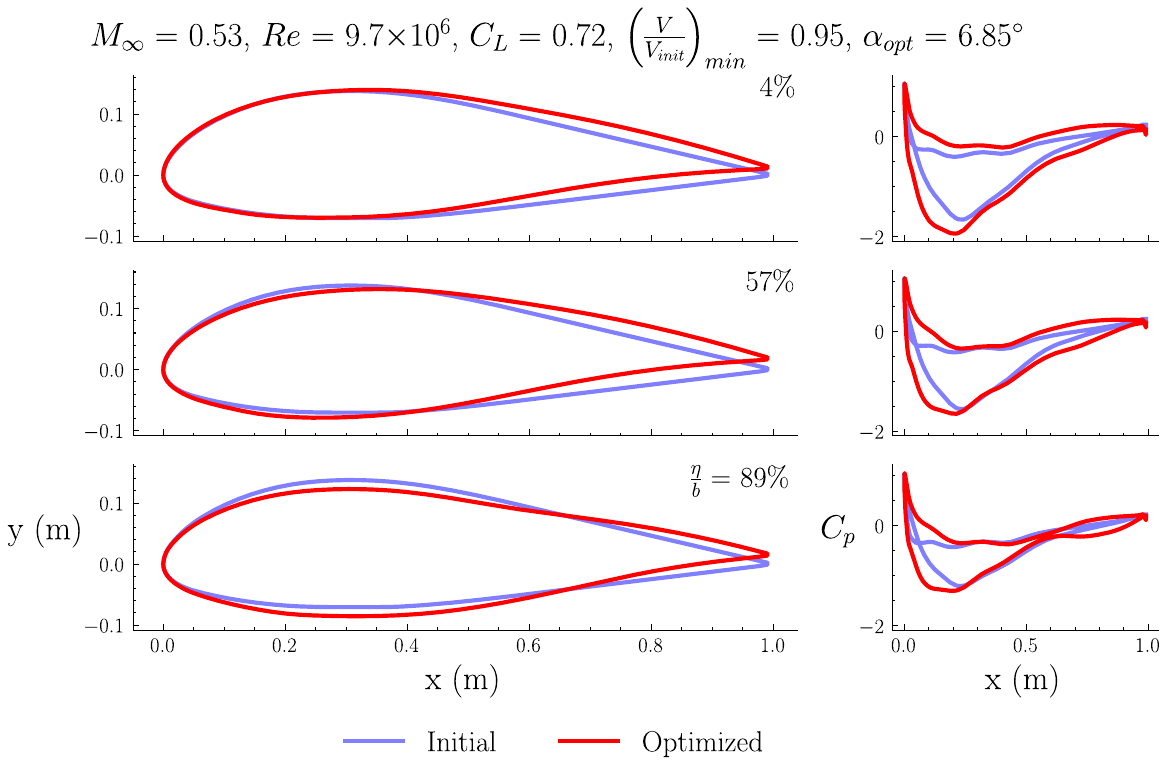}
        \caption{Sample c}
        \label{fig:sample_c_opt_init}
    \end{subfigure}
    \hfill
    \begin{subfigure}[b]{0.495\textwidth}
        \centering
        \includeinkscape[width=\textwidth]{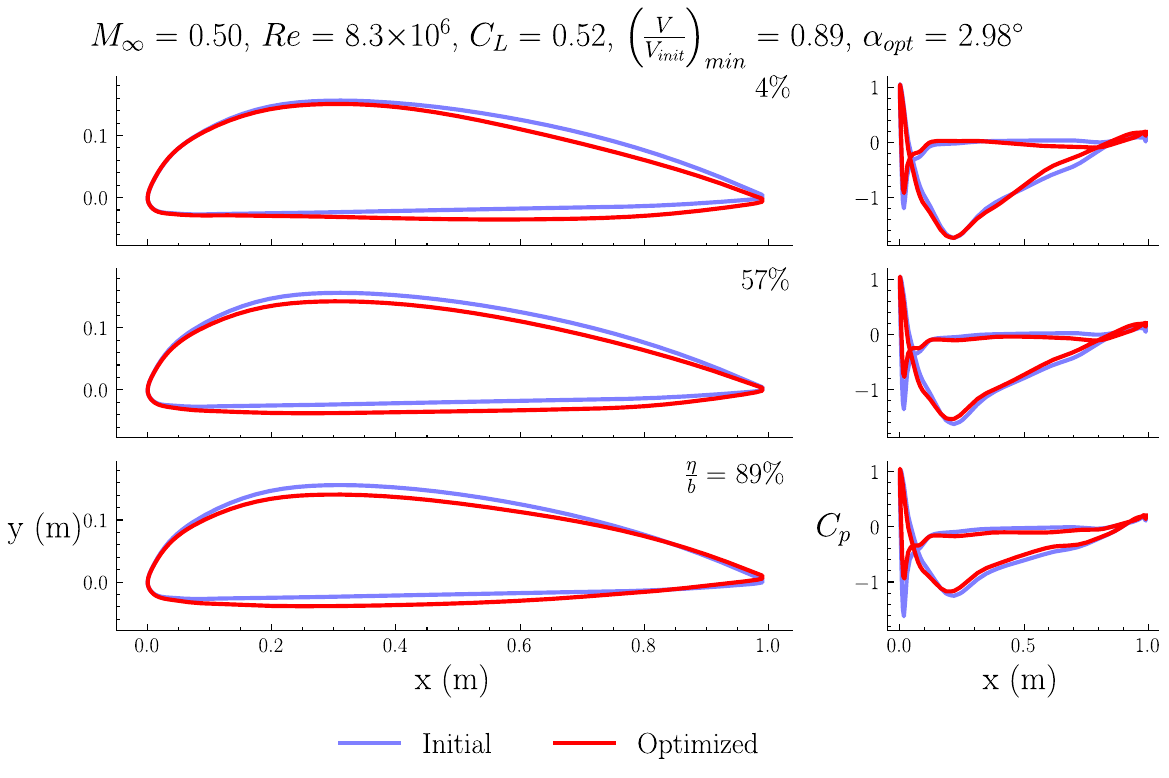}
        \caption{Sample d}
        \label{fig:sample_d_opt_init}
    \end{subfigure}
        \caption{Random samples: optimized, initial geometries and surface pressures}
        \label{fig:sample_opt_init}
\end{figure}

In Figure~\ref{fig:sample_opt_init}, we present a random subset of optimized designs cross sections, at 3 spanwise locations, and their corresponding pressure distributions, from the dataset. 

\begin{figure}[H]\centering
\includeinkscape[width=0.80\textwidth]{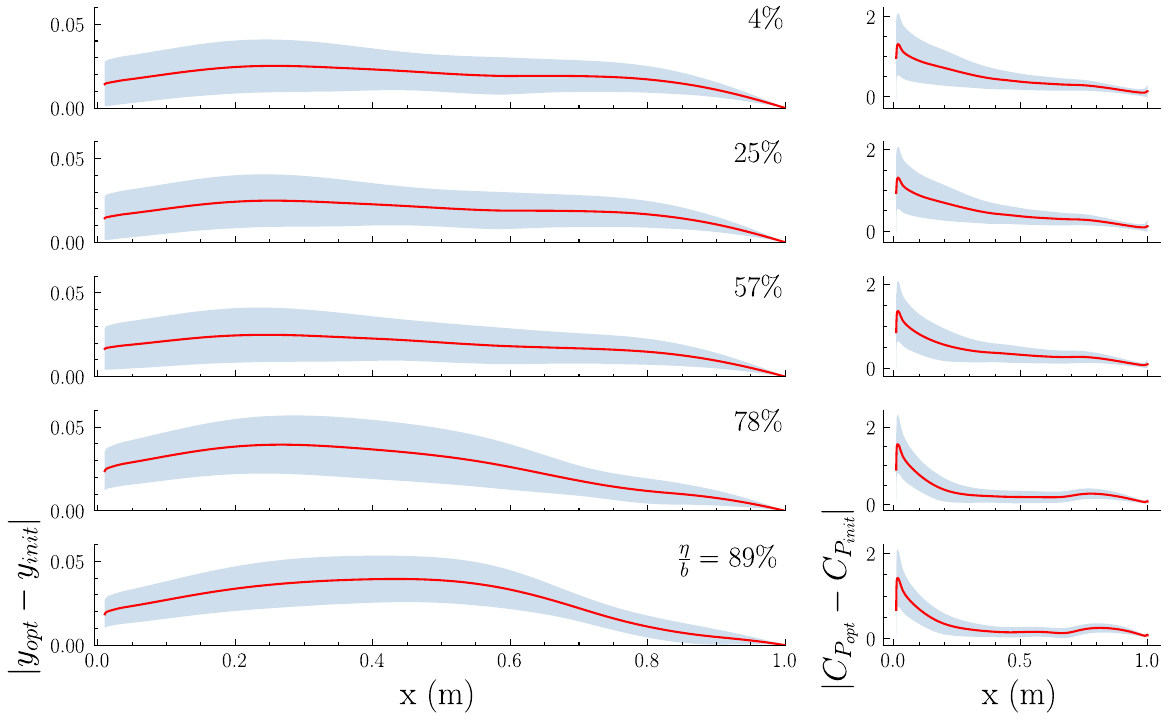}
\caption{Aggregate optimized shape and pressure differences}
\label{fig:agg_opt_diff}
\end{figure}

These samples show that shape variation is relatively minor, especially near the root of the wing, although the differences in pressure distributions can also be significant despite this. Figure~\ref{fig:agg_opt_diff} shows the aggregate magnitude of the shape changes, for a selected number of slices and percentages along the half-span ($\frac{\eta}{b}$), across the dataset. 

It appears that most optimization-related shape changes occur in the last 20\% of the span closest to the tip. Intuitively, a bias towards the tip is to be expected since the vortices due to 3D effects are strongest at the wingtips~\cite{Anderson2011Fundamentals}. This indicates that the inboard sections are closer to 2D-optimized designs already, so less correction is needed. It is also observed that the greatest difference in shape appears to occur closer to the leading edge of the airfoil. Further down the span, the greatest difference in shape moves closer to the center, while the difference at the trailing edge slopes down more dramatically.  

\begin{figure}[H]
    \centering
    \begin{subfigure}[b]{0.48\textwidth}
        \centering
        \includegraphics[width=\textwidth]{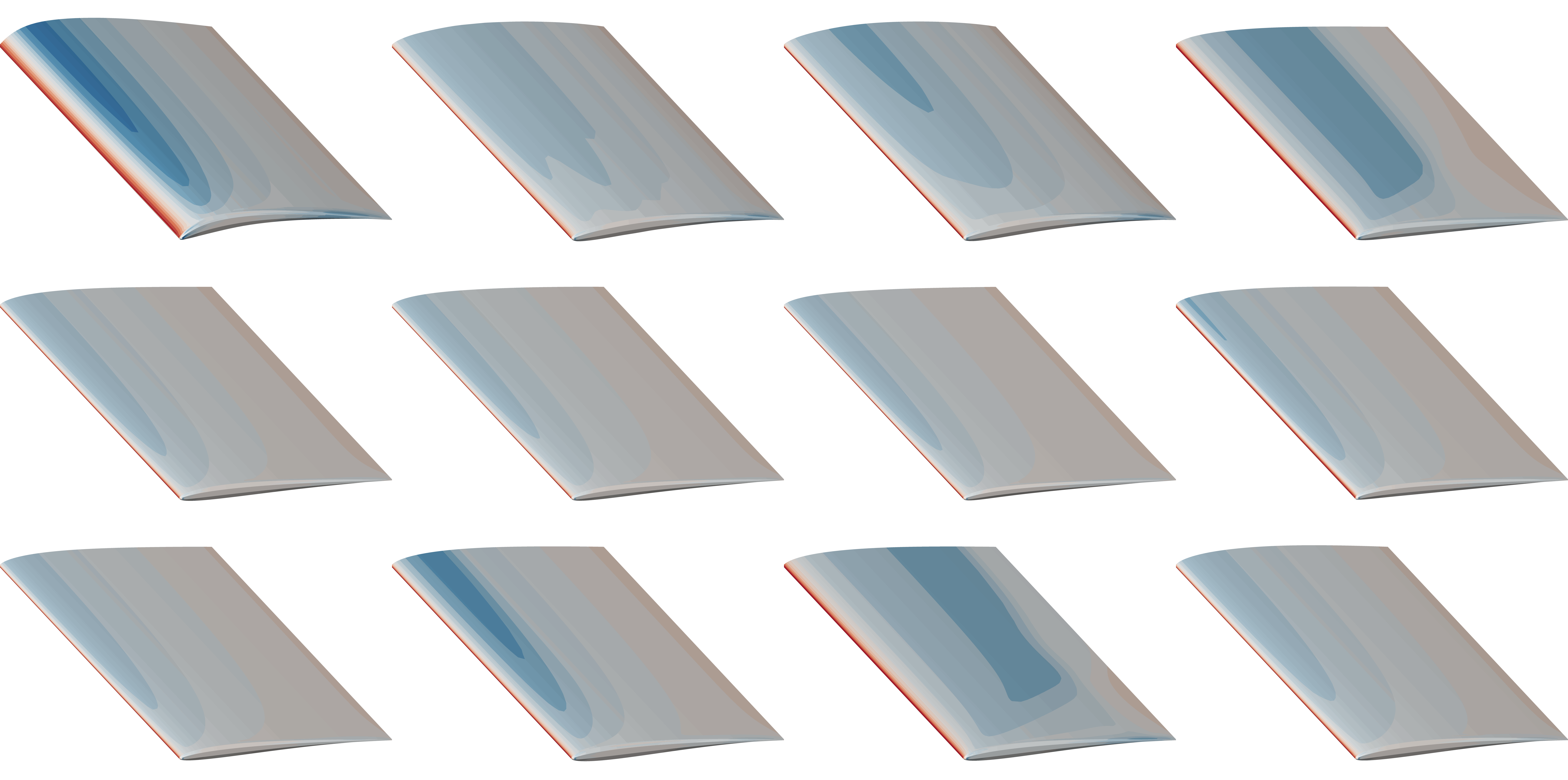}
        \caption{Initial wings}
        \label{fig:cp_init}
    \end{subfigure}
    \hfill
    \begin{subfigure}[b]{0.48\textwidth}
        \centering
        \includegraphics[width=\textwidth]{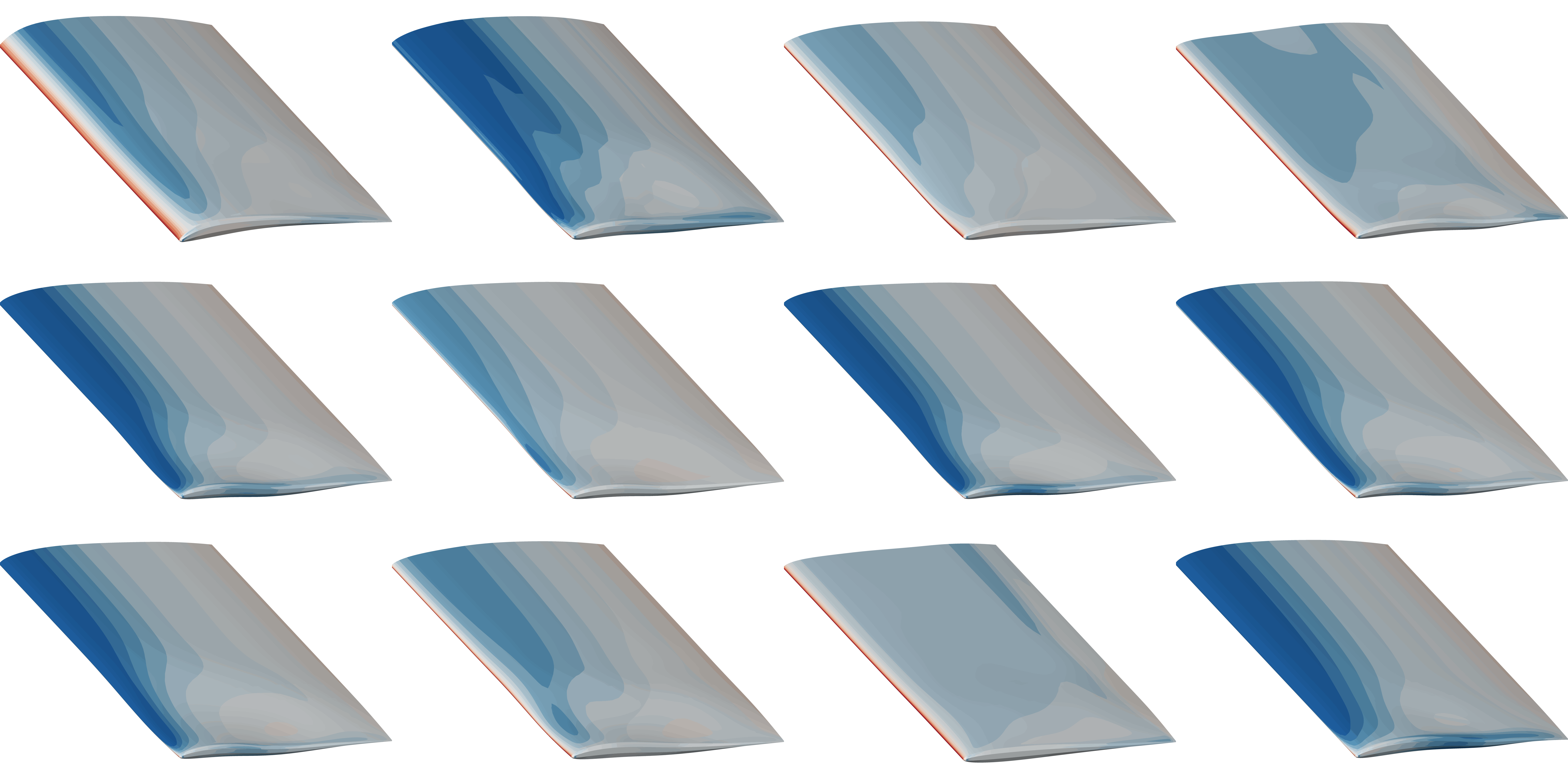}
        \caption{Optimized wings}
        \label{fig:cp_opt}
    \end{subfigure}
        \caption{Initial and corresponding optimized 3D pressure distributions from the dataset}
        \smallskip
        \small
        Red and blue contours correspond to higher and lower pressure regions, respectively.
        \label{fig:cp_opt_init}
\end{figure}

Finally, as a visual reference, we present isometric views of the 3D pressure coefficient distribution for initial and optimized wings from the dataset in Figure~\ref{fig:cp_opt_init}.

\subsection{2D-3D Optimization Comparison}

\begin{figure}[H]
    \centering
    \begin{subfigure}[b]{0.495\textwidth}
        \centering
        \includeinkscape[width=\textwidth]{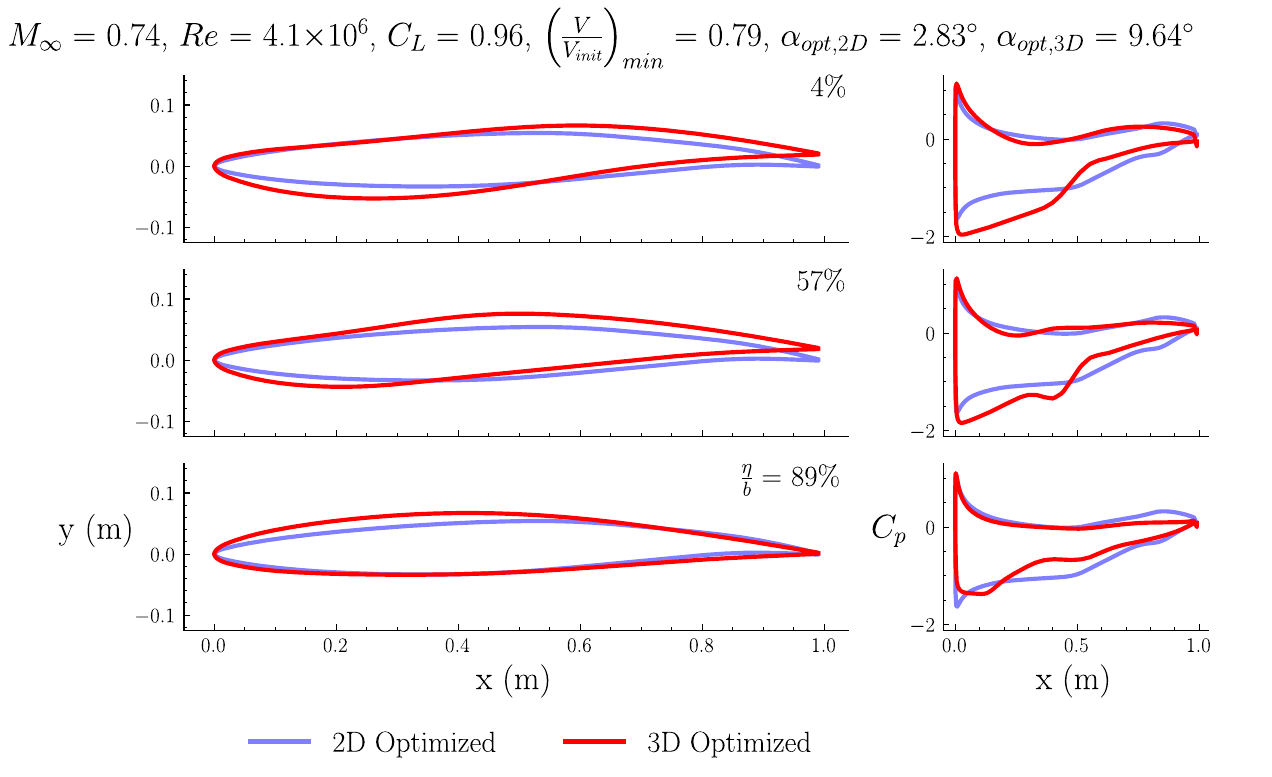}
        \caption{Sample a}
        \label{fig:sample_a_opt_2d3d}
    \end{subfigure}
    \hfill
    \begin{subfigure}[b]{0.495\textwidth}
        \centering
        \includeinkscape[width=\textwidth]{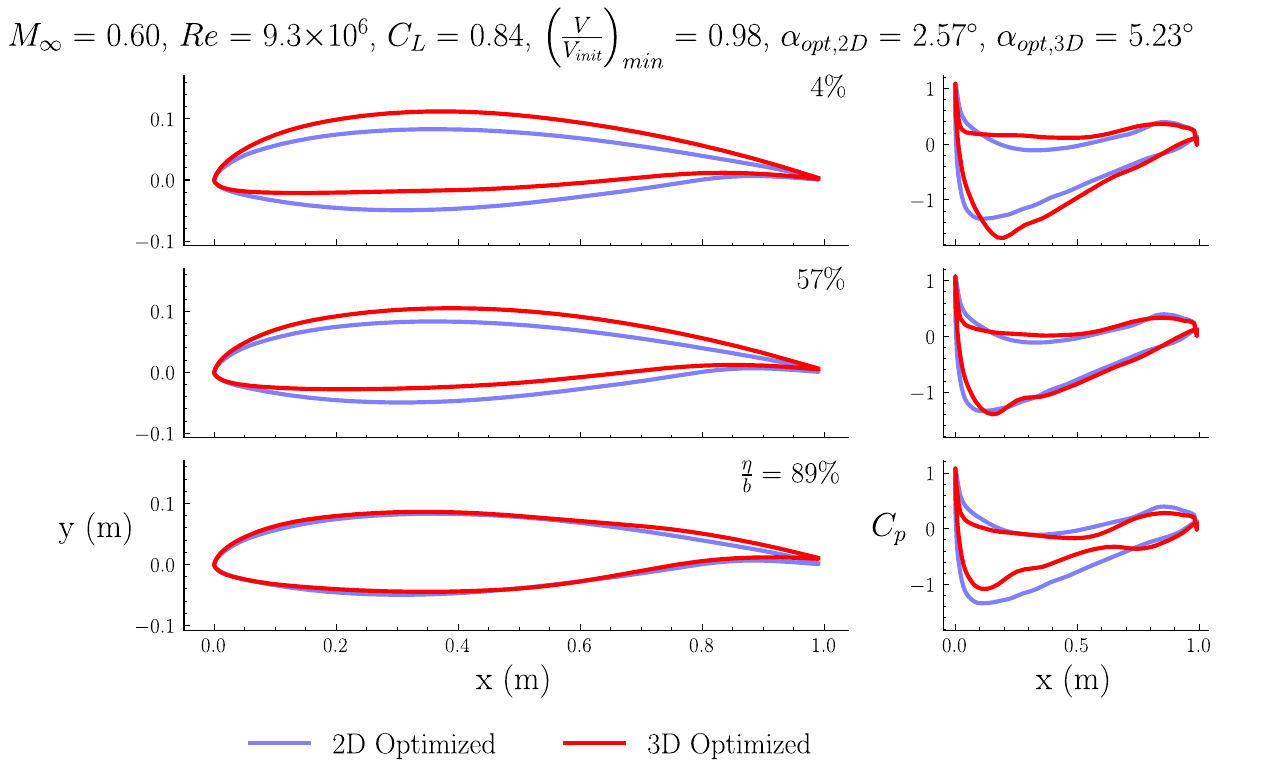}
        \caption{Sample b}
        \label{fig:sample_b_opt_2d3d}
    \end{subfigure}
    \hfill
    \begin{subfigure}[b]{0.495\textwidth}
        \centering
        \includeinkscape[width=\textwidth]{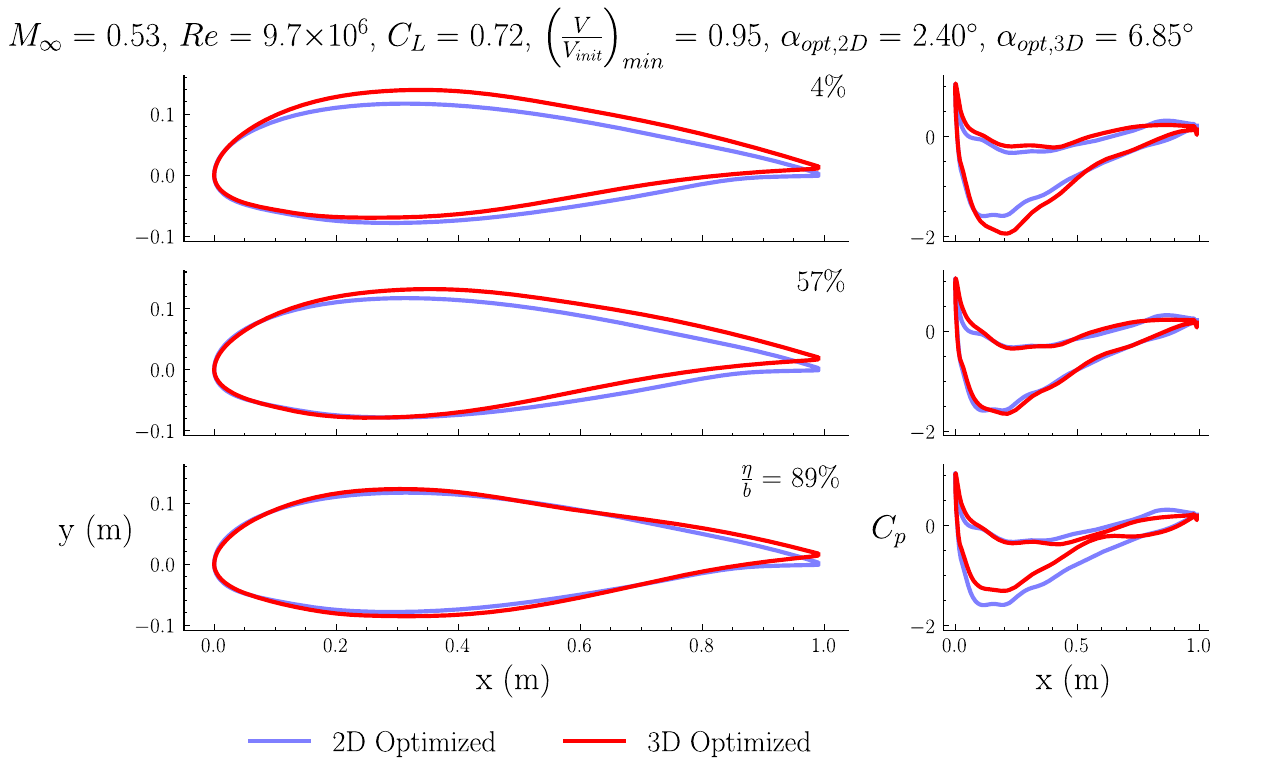}
        \caption{Sample c}
        \label{fig:sample_c_opt_2d3d}
    \end{subfigure}
    \hfill
    \begin{subfigure}[b]{0.495\textwidth}
        \centering
        \includeinkscape[width=\textwidth]{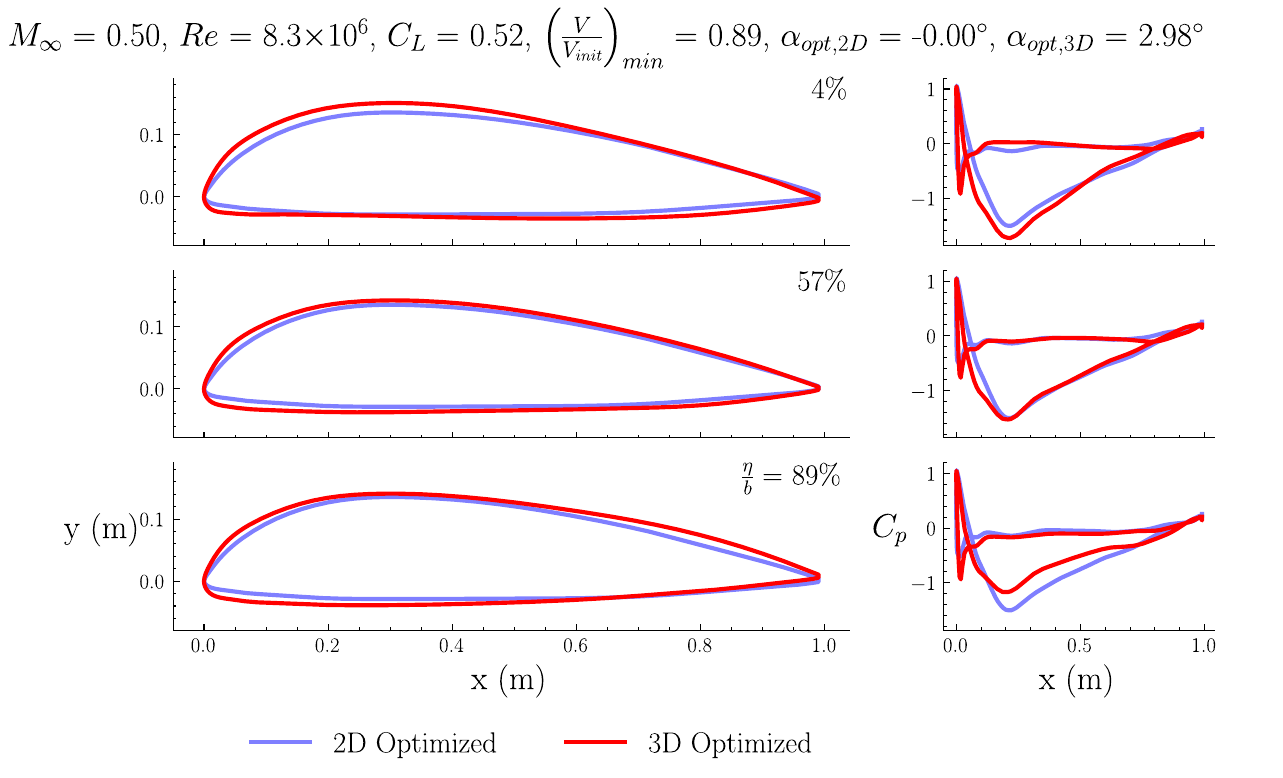}
        \caption{Sample d}
        \label{fig:sample_d_opt_2d3d}
    \end{subfigure}
        \caption{Random samples: optimized 2D, 3D geometries and surface pressures}
        \label{fig:sample_opt_2d3d}
\end{figure}

Here, we present some basic visual comparisons between the 2D and 3D optimized designs. Figure~\ref{fig:sample_opt_2d3d} plots the optimized 2D airfoil against 3D optimized wing slices for random samples from the dataset. Note that the 2D optimized design is constant and is plotted multiple times at different 3D span locations purely as a visual reference. 

\begin{figure}[H]\centering
\includeinkscape[width=0.80\textwidth]{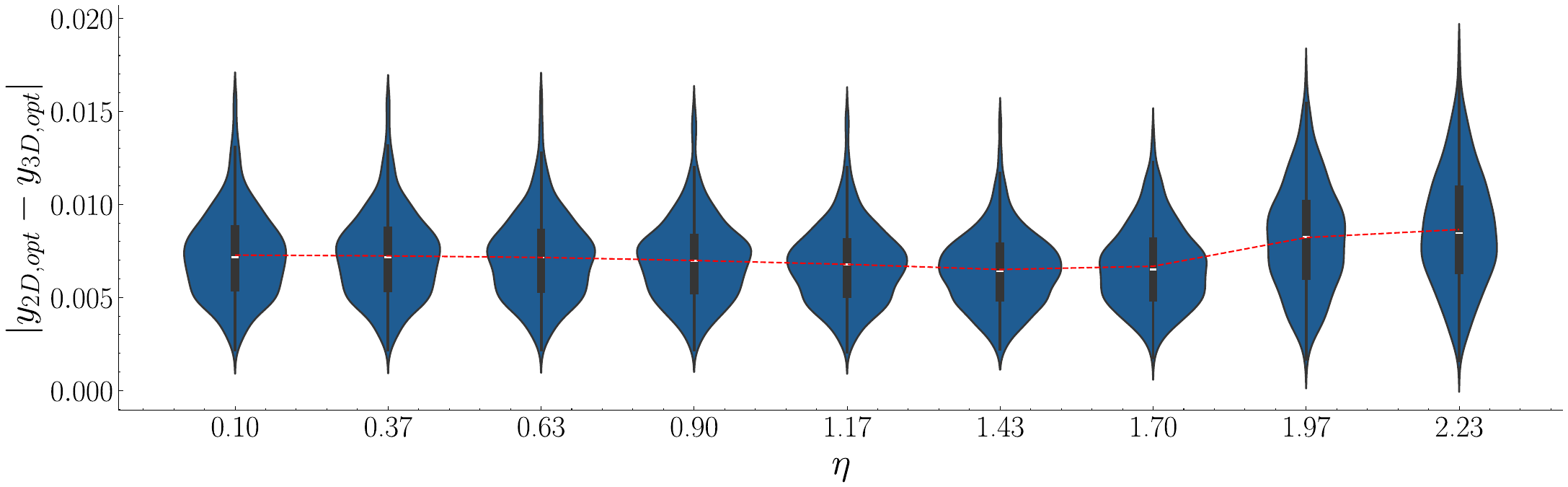}
\caption{Aggregate 2D-3D shape and pressure differences}
\label{fig:agg_2d3d_opt_shape}
\end{figure}

Figure~\ref{fig:agg_2d3d_opt_shape} presents the aggregate difference in shape for the optimized 2D and 3D designs. It is notable that in Figure~\ref{fig:sample_opt_2d3d} the difference in the optimized designs appears to be smallest at the tip in this random sample. However, according to the aggregate difference in Figure~\ref{fig:agg_2d3d_opt_shape}, the greatest mean difference (plotted as the red dotted line) appears to occur closest to the wingtip. This is most likely due to random sampling, but also reinforces the fact that the distribution of shape differences is also much wider closer to the tip. 

\begin{figure}[H]
    \centering
    \begin{subfigure}[b]{0.495\textwidth}
        \centering
        \includeinkscape[width=\textwidth]{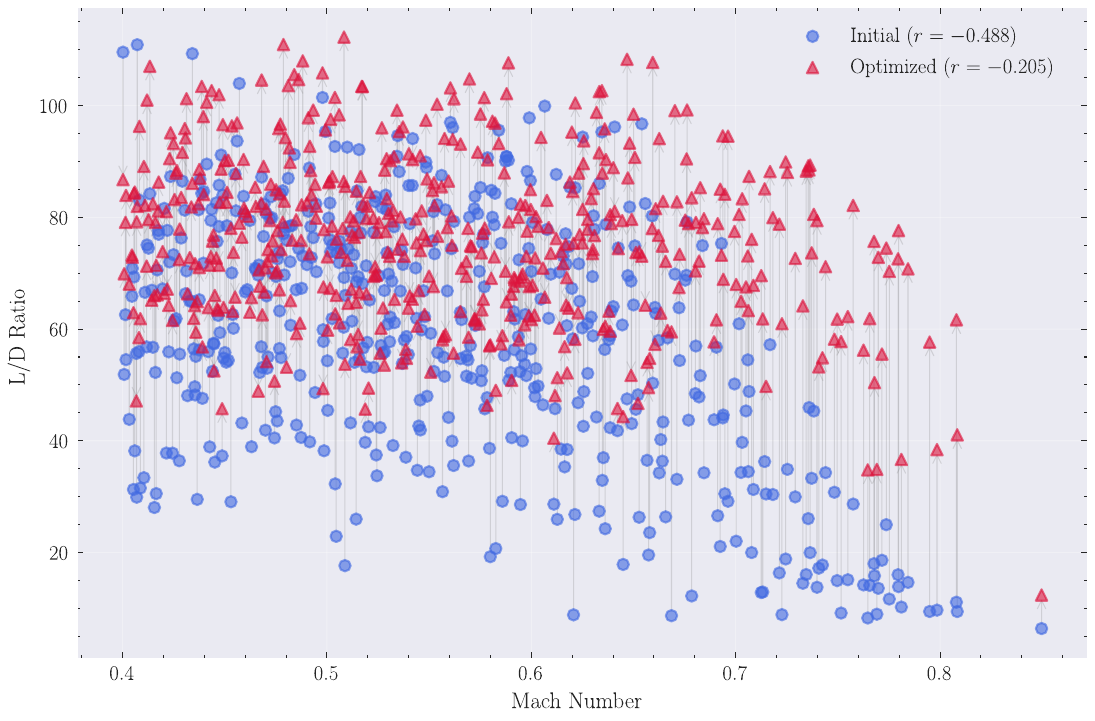}
        \caption{2D Performance Distribution: Mach Variation}
        \label{fig:ld_vs_mach_2d}
    \end{subfigure}
    \hfill
    \begin{subfigure}[b]{0.495\textwidth}
        \centering
        \includeinkscape[width=\textwidth]{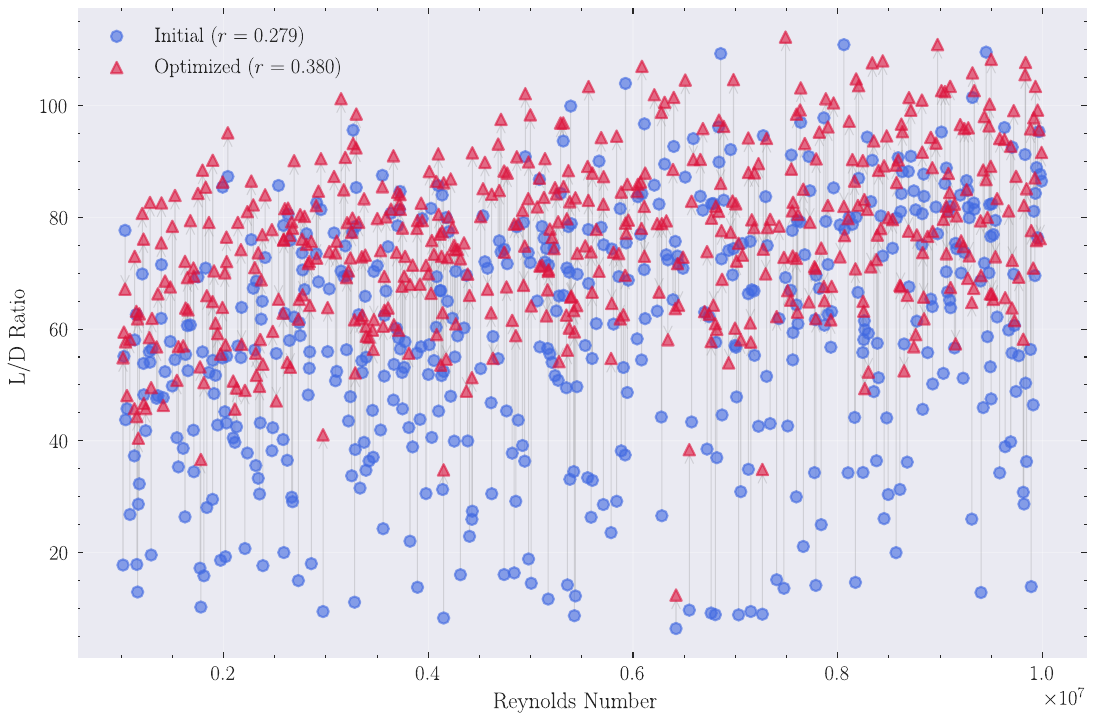}
        \caption{2D Performance Distribution: Reynolds Number}
        \label{fig:ld_vs_reynolds_2d}
    \end{subfigure}
    \hfill
    \begin{subfigure}[b]{0.495\textwidth}
        \centering
        \includeinkscape[width=\textwidth]{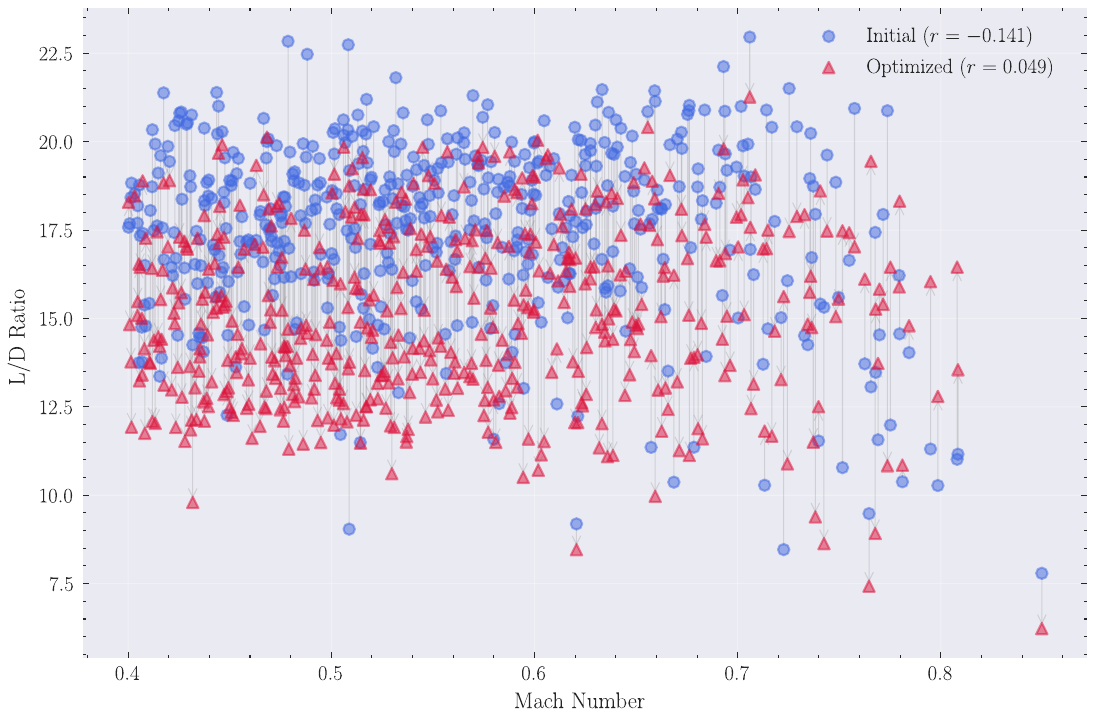}
        \caption{3D Performance Distribution: Mach Variation}
        \label{fig:ld_vs_mach_3d}
    \end{subfigure}
    \hfill
    \begin{subfigure}[b]{0.495\textwidth}
        \centering
        \includeinkscape[width=\textwidth]{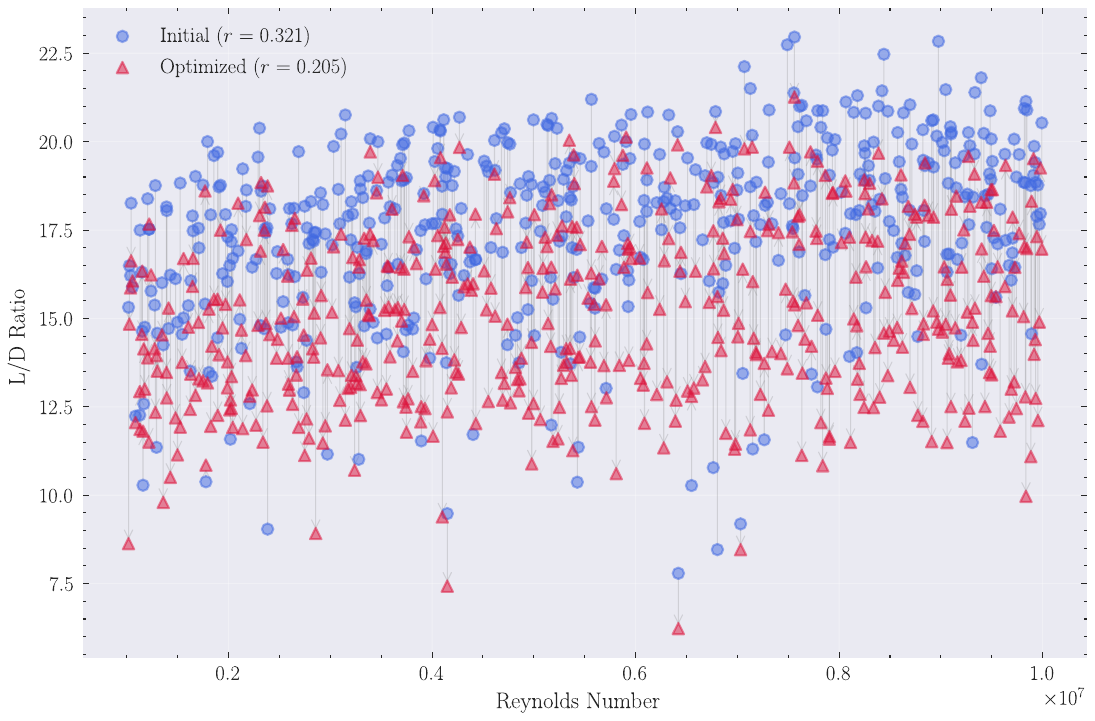}
        \caption{3D Performance Distribution: Reynolds Number}
        \label{fig:ld_vs_reynolds_3d}
    \end{subfigure}
        \caption{Dataset Performance Distributions}
        \label{fig:perf_init_opt}
\end{figure}

In addition to the aggregate shape differences, we attempt to visualize the distributional change in performance space between the initial and optimized designs as a function of input flow conditions. Figure~\ref{fig:perf_init_opt} presents the lift to drag ratio distributions as a function of Mach and Reynolds numbers for both the 2D and 3D datasets. It should be noted that in order to create a fair comparison, only cases where the initial design was able to simulate were included (for the cases where the initial design failed to simulate, the optimized designs were paired with the first successful intermediate design instead). It should also be noted that $\frac{L}{D}$ is plotted for ease of visualization (due to larger differences in drag values) in the optimized cases, as $C_{L}$ is a sampled constraint, and is thus uncorrelated to the input flow conditions.

In absolute terms, the 2D optimized wing designs appear to exhibit increased performance, while the 3D optimized designs exhibit a reduced $\frac{L}{D}$ ratio relative to the initial design performance. This discrepancy may come from 3-dimensional effects, such as induced drag (a 3D-specific effect), which can make up a non-negligible fraction of the total drag in the case of finite wings. As a result, for the same given $C_{L}$ constraint, the optimizer contends with an added induced drag penalty, which cannot be offset by local shape variation alone, explaining the resultant performance decrease. 

Additionally, the correlation between flow conditions and performance differs by fidelity. There is a somewhat direct relationship between Mach and $\frac{L}{D}$ in both datasets although the correlation does weaken for the 3D optimized designs, while conversely strengthening for the 2D designs. This suggests additional variability resulting from 3D effects could obfuscate Mach number dependence. For the Reynolds number, we observe an indirect, and weaker correlation with $\frac{L}{D}$ across all cases. The one exception is for the 2D cases, where there is a noticeable performance drop after a Mach number of around 0.7. This could indicate the phenomena of transonic drag rise (the increase in drag observed in the transonic regime due to shock formation), and that this phenomena is more pronounced in the 2D cases, perhaps due to additional complexities introduced in 3D. 

\subsection{Dimensional Analysis}

\begin{figure}[H]
    \centering
    \begin{subfigure}[b]{0.45\textwidth}
        \centering
        \includeinkscape[width=\textwidth]{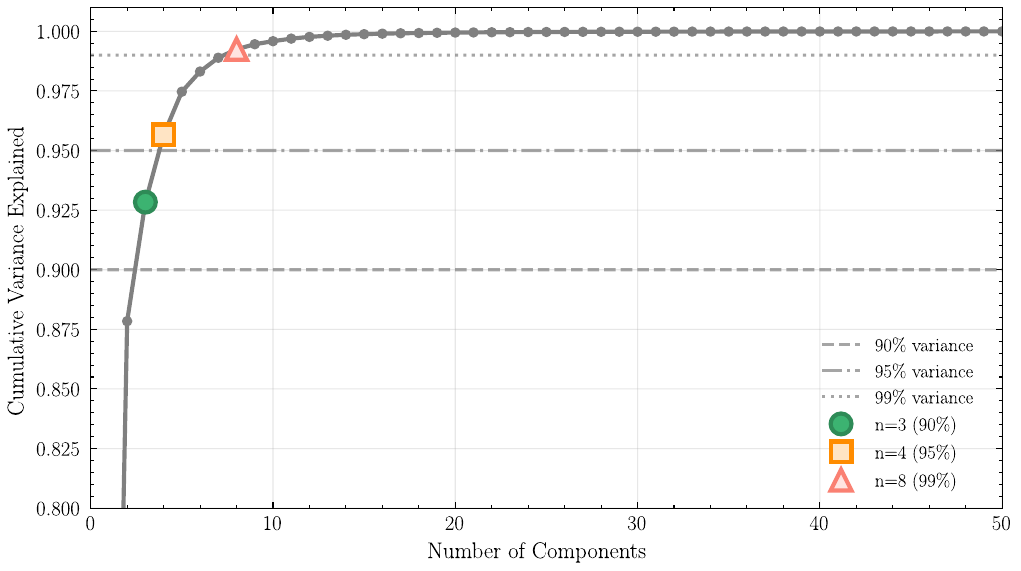}
        \caption{2D Initial Airfoils}
        \label{fig:pca_shape_2d_init}
    \end{subfigure}
    \hfill
    \begin{subfigure}[b]{0.45\textwidth}
        \centering
        \includeinkscape[width=\textwidth]{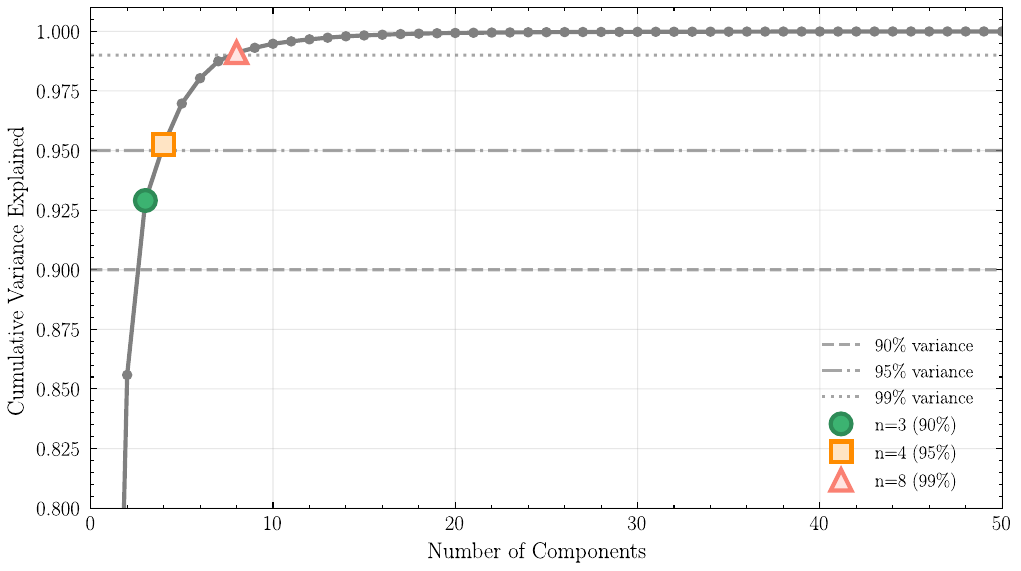}
        \caption{2D Optimized Airfoils}
        \label{fig:pca_shape_2d_opt}
    \end{subfigure}
    \hfill
    \begin{subfigure}[b]{0.45\textwidth}
        \centering
        \includeinkscape[width=\textwidth]{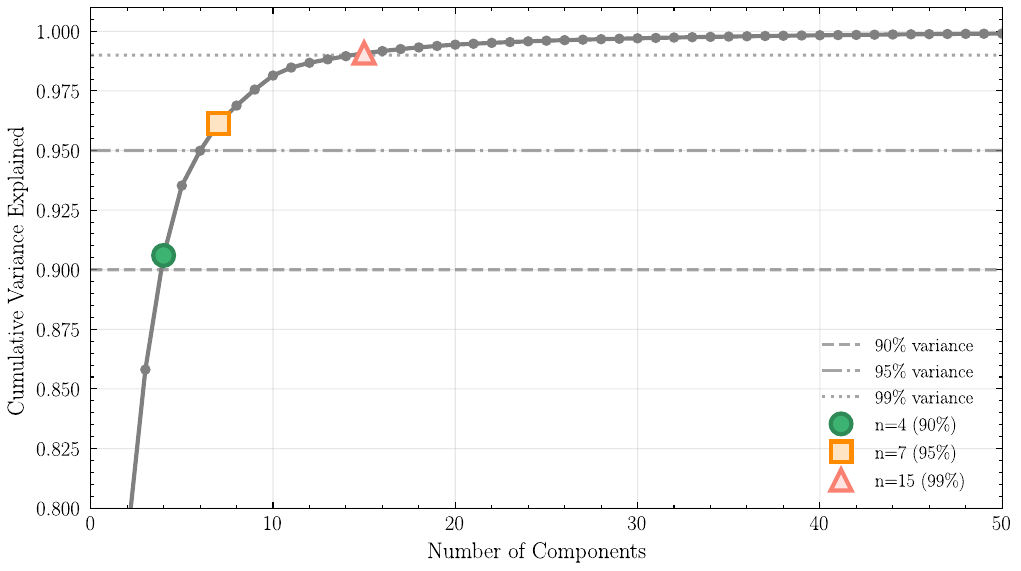}
        \caption{3D Optimized Wings}
        \label{fig:pca_shape_3d_opt}
    \end{subfigure}
        \caption{Cumulative Variance Explained: Dataset Geometries}
        \label{fig:pca_shape}
\end{figure}

In order to study the complexity and diversity, we conduct a simple dimensionality analysis of the OptiWing3D dataset. In order to obtain an estimate of intrinsic dimensionality, we employ a principal component analysis (PCA). We present the cumulative explained variance as a function of the number of principal components for the 3D and 2D initial and optimized geometries in Figure~\ref{fig:pca_shape}. Note that for the purposes of this analysis (and throughout this paper), we restrict ourselves to 9 linearly-spaced slices along the span (not including the wingtip) for the 3D cases. Additionally, since the initial 3D wings are simply extrusions of the 2D initial airfoil geometries, we forgo any redundant analysis of the initial 3D geometries. 

Achieving $n=99\%$ cumulative explained variance requires around 8 components for both the initial and optimized airfoil geometries, indicating that complexity stays roughly the same in design space. For the 3D optimized designs, the dimensionality is almost doubled, with 15 dimensions required to achieve the same cumulative explained variance, indicating greater geometric complexity introduced by 3D optimization.

\begin{figure}[H]
    \centering
    \begin{subfigure}[b]{0.45\textwidth}
        \centering
        \includeinkscape[width=\textwidth]{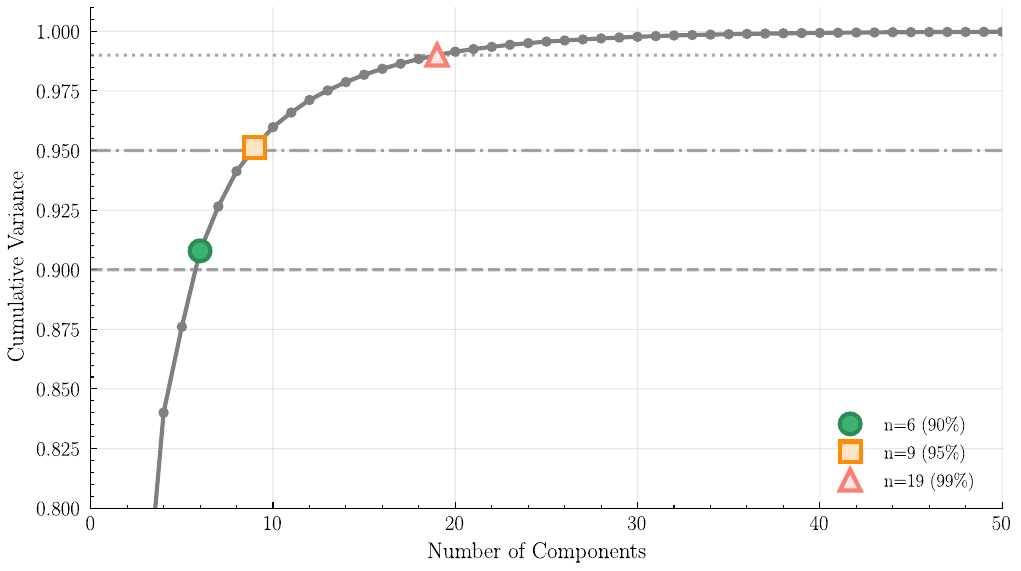}
        \caption{2D Initial Airfoils}
        \label{fig:pca_pressure_2d_init}
    \end{subfigure}
    \hfill
    \begin{subfigure}[b]{0.45\textwidth}
        \centering
        \includeinkscape[width=\textwidth]{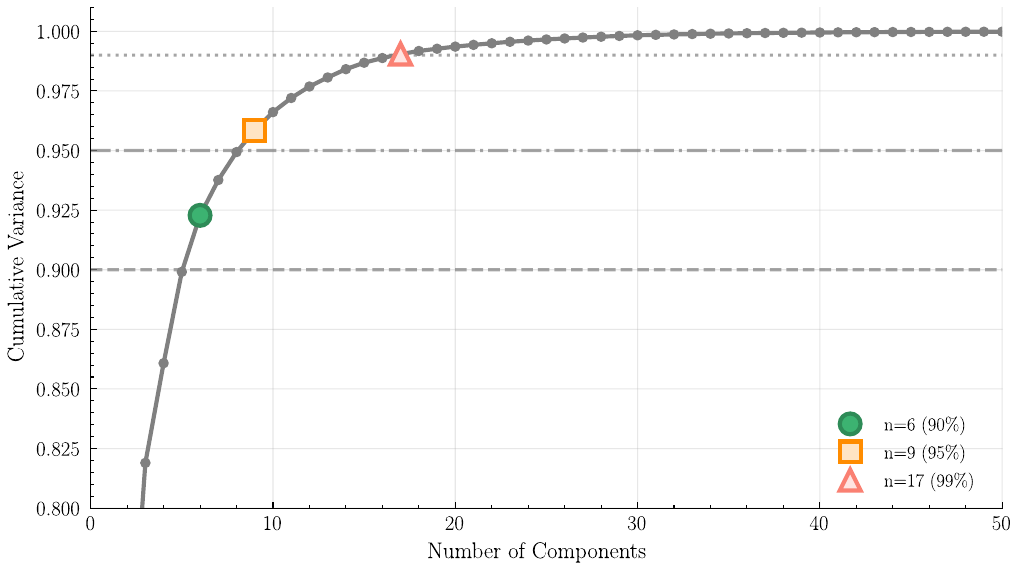}
        \caption{2D Optimized Airfoils}
        \label{fig:pca_pressure_2d_opt}
    \end{subfigure}
    \hfill
    \begin{subfigure}[b]{0.45\textwidth}
        \centering
        \includeinkscape[width=\textwidth]{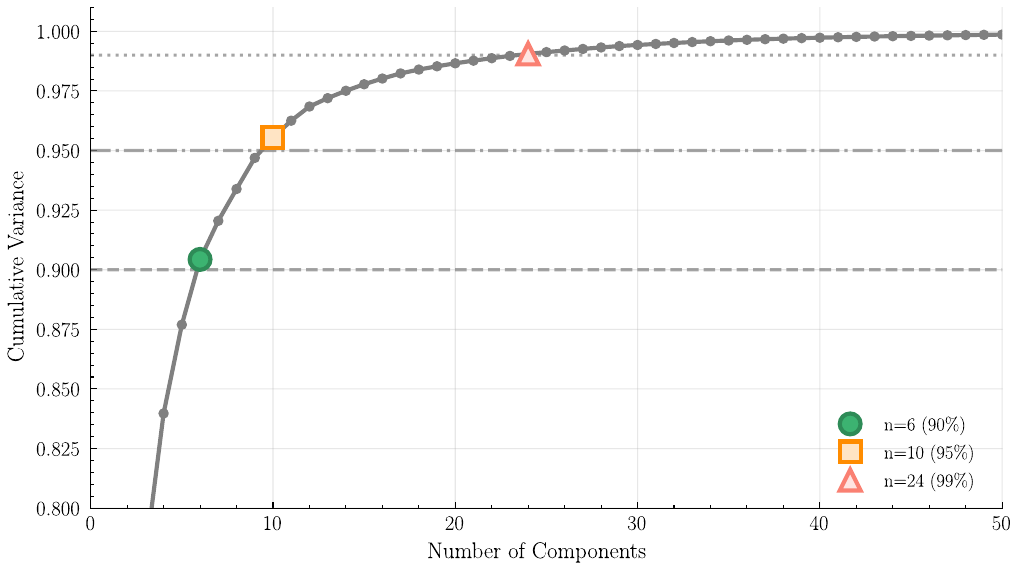}
        \caption{3D Initial Wings}
        \label{fig:pca_pressure_3d_init}
    \end{subfigure}
    \hfill
    \begin{subfigure}[b]{0.45\textwidth}
        \centering
        \includeinkscape[width=\textwidth]{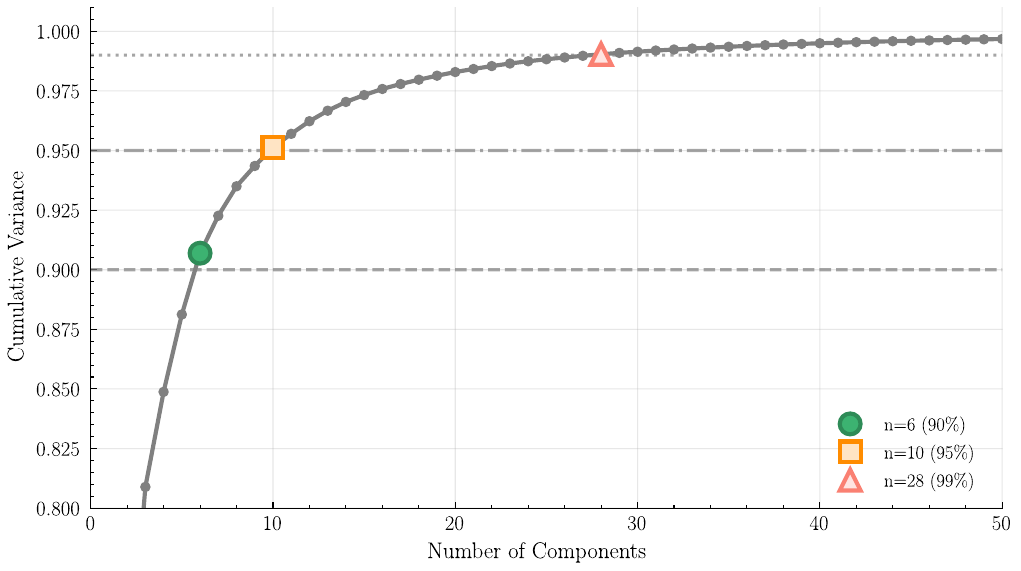}
        \caption{3D Optimized Wings}
        \label{fig:pca_pressure_3d_opt}
    \end{subfigure}
        \caption{Cumulative Variance Explained: Surface Pressure Distributions}
        \label{fig:pca_pressure}
\end{figure}

Although variance in geometry may be helpful in constructing a rough estimate of the underlying dimensionality of the dataset, it may be biased, as certain performance-space features may be unduly sensitive to certain geometric features, even if those features may only explain a relatively small amount of the dataset's variance. To address this, we also examine the dimensionality of the performance space, specifically, the dimensionality of the surface pressure distributions. We present the explained variance of the pressure distributions in Figure~\ref{fig:pca_pressure}. 

In performance space, we see that the 2D pressure distributions require roughly 19 components to achieve $n=99$\% cumulative explained variance for the initial geometries, and 17 for the optimized geometries. It is notable, that this is a decrease in complexity from initial to optimized. For the 3D pressure distributions, the required number of components increases from 24 to 28, from initial to optimized. Although these dimensionality estimates do not give a relation to geometric sensitivities, compared to the geometric dimensionality estimates, they are higher across the board for both the initial and optimized designs. As such, it may be reasonable to assume that the number of dimensions required to faithfully reconstruct performant airfoil and wing geometries will be higher than the geometric-only estimates imply.

\section{Generative Model Setup}

\subsection{Overview}

In order to create an accurate baseline generative model, we have chosen to implement a conditional latent denoising diffusion model (DDM) with the capability to predict optimized wing designs, $W_{\text{opt}}$, and their angle of attack, $\alpha$. We again note that during optimization, airfoil sections were permitted to shift in the y-axis (span-perpendicular direction), effectively allowing small dihedral shifts, $\eta_{y}$. Because of these shifts, we choose to encode the unshifted coordinates, $W_{unshifted}$, into an autoencoder-learned latent space (described in this section) for smoother, more accurate coordinate embeddings. Our model predicts $\eta_{y}$ and adds it (span-wise) internally to the unshifted wing cross sections, $W_{unshifted}$, i.e: $W_{opt}$ as $W_{opt}= W_{unshifted} + \eta_{y,opt}$. In addition, in our dataset, we use 9 discrete, linearly spaced airfoil sections along the span to be representative of the entire wing geometry. 

The model takes as inputs the flow conditions (Mach number, $M$ and Reynolds number, $Re$), the constraints (lift coefficient, $C_{L}$, and minimum allowable volume, $V_{min}$), and the initial extruded airfoil cross section, $A_{0}$.

We choose a DDM for the 3D wing generation task, principally because it has been shown to be effective in the generation of 2D airfoil designs, both in our previous work, where it was shown to have beaten performant optimal-transport based approaches, and elsewhere in literature~\cite{diniz2024optimizing, graves2024airfoil, yonekura2024preliminary}. 

\subsection{Diffusion Model Background}

Although interested readers may reference previous work on DDMs (see~\cite{diniz2024optimizing}), we will briefly summarize how DDMs work here:

Diffusion models are a class of generative models that learn a data distribution $p(\mathbf{x})$ through a noise-based iterative process. For complex, high-dimensional data such as wing geometries, directly modeling $p(\mathbf{x})$ is computationally intensive. Instead, diffusion models introduce a latent variable $\mathbf{z}$ and maximize a tractable Evidence Lower Bound (ELBO) on the log-likelihood~\cite{luo2022understanding}:
\begin{equation}
\log p(\mathbf{x}) \geq \mathbb{E}_{q_{\boldsymbol{\theta}}(\mathbf{z} | \mathbf{x})} \left[\log \frac{p(\mathbf{x}, \mathbf{z})}{q_{\boldsymbol{\theta}}(\mathbf{z} | \mathbf{x})}\right]
\end{equation}
where $q_{\boldsymbol{\theta}}(\mathbf{z} | \mathbf{x})$ is a variational distribution with tunable weights $\boldsymbol{\theta}$.

Diffusion models operate by two main processes. The \emph{forward process}, used in training, progressively adds Gaussian noise to data over $T$ timesteps according to a variance schedule $\beta_t$:
\begin{equation}
q(\mathbf{x_t} | \mathbf{x_{t-1}}) = \mathcal{N}(\mathbf{x_t}; \sqrt{\alpha_t} \mathbf{x_{t-1}}, 1-\alpha_t)
\end{equation}
where $\alpha_t = 1 - \beta_t$ and $t \in [0, T]$. At $t=0$, $\mathbf{x_0}$ represents the original data, while as $t \to T$, the data approaches an isotropic Gaussian distribution. In this work, we employ a linear variance schedule. 

The \emph{reverse process} learns to denoise, generating samples by iteratively removing noise:
\begin{equation}
p_{\boldsymbol{\theta}}(\mathbf{x_{t-1}} | \mathbf{x_t}) = \mathcal{N}(\mathbf{x_{t-1}}; \boldsymbol{\mu}_{\boldsymbol{\theta}}(\mathbf{x_t}, t), \boldsymbol{\sigma}_{\boldsymbol{\theta}}(\mathbf{x_t}, t))
\end{equation}
where the mean $\boldsymbol{\mu}_{\boldsymbol{\theta}}$ and variance $\boldsymbol{\sigma}_{\boldsymbol{\theta}}$ are parameterized by a neural network. In practice, the model predicts the noise, $\epsilon$, added to the data, and the ELBO objective simplifies to minimizing the Mean Squared Error (MSE) between predicted and actual noise~\cite{ho2020denoising}. For a comprehensive derivation and detailed treatment of diffusion models, we refer readers to~\cite{luo2022understanding, ho2020denoising}.

\subsection{Latent Space Encoding}

\begin{figure}[H]\centering
{\includeinkscape[width=.50\linewidth]{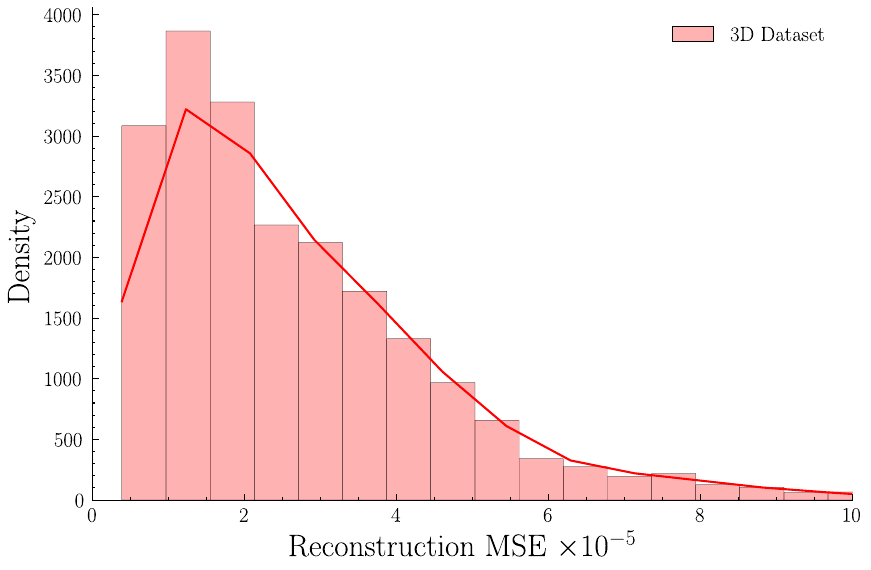}}
\caption{MSE Distribution of BAE 3D Slice Reconstructions}
\smallskip
\small
The solid line represents Gaussian kernel density estimates for over the entire 3D dataset. The shaded regions are histogram bins. 
\label{fig:BezierErrorDist}
\end{figure}

In practice, our DDM implementation works with a \emph{latent} representation of the wing coordinates. In this case, we follow previous work~\cite{diniz2024optimizing}, and train an autoencoder model to transform each airfoil's cross sectional coordinates into a latent space. The Bezier autoencoder's (BAE) encoder component, $\varepsilon_{z}$, encodes 2D airfoil coordinates, $A$, into a Bezier control point and weight vector, $Z_{A}$ (i.e $x$, $y$, and $w$). In this work, 30 control points and weights are used, as we found this allows for reasonable reconstruction accuracy. During training, the decoder component of the BAE, $D_{z}$, attempts to decode the latent control point and weight vector back into coordinate space. As in previous work, encoding into a Bezier control point and weight space ensures that the design representation in smooth in the coordinate space, which is crucial for performant design synthesis. In this work, the BAE is trained solely on the 2D data, and is given access to the first 25\% of the intermediate designs in each optimization trajectory, or around ~20k perturbed designs in total. We chose the first 25\% of intermediate designs, as prior work has shown this fraction contains the majority of shape variation~\cite{diniz2024optimizing}.  To validate our BAE implementation, we present the error distribution over the entire 3D dataset in Figure~\ref{fig:BezierErrorDist}. 

In general, the error produced by the BAE reconstructions is low, although we accept that given the relatively long tail of the error distribution, further improvements may be needed to improve outlier performance.

\subsection{Diffusion Model Implementation}

\begin{figure}[H]\centering
\includeinkscape[width=0.80\textwidth]{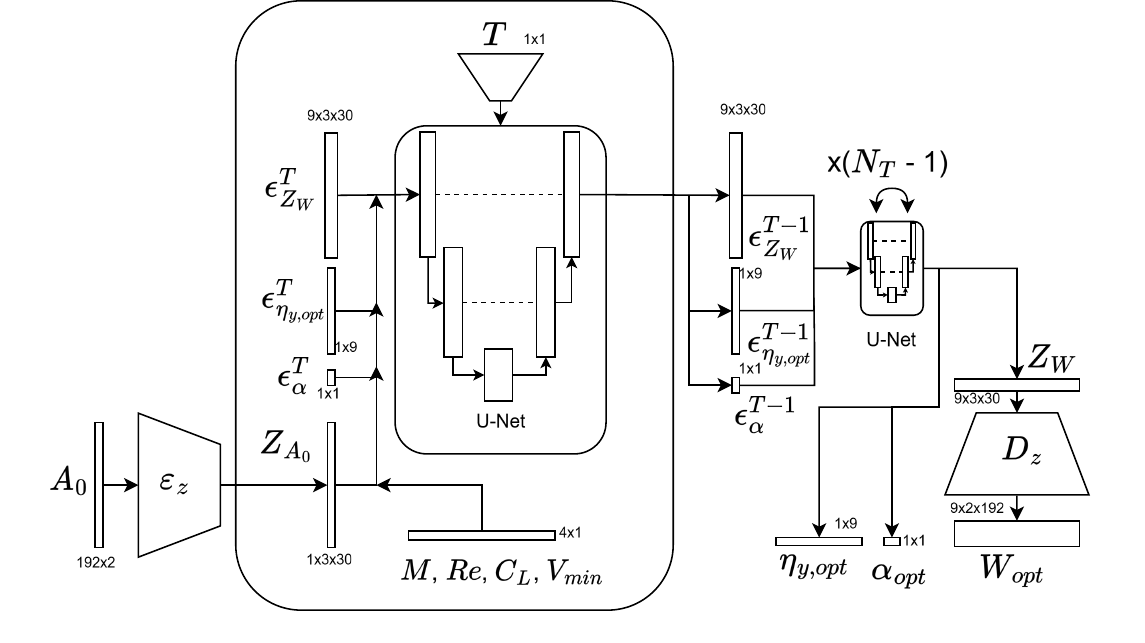}
\caption{Wing Diffusion Model Architecture}
\label{fig:ddm_architecture}
\end{figure}

A high-level diagram of the diffusion model architecture is presented in Figure~\ref{fig:ddm_architecture}. 

In addition to the flow conditions constraints, and Bezier encoded initial airfoil cross section, $Z_{A_{0}}$, internally, the model takes in noised versions, at a given timestep $T$, of the Bezier encoded optimized wings $\epsilon_{Z_{W}}$, the required span offsets, $\epsilon_{\eta_{y}}$, and the angle of attack, $\epsilon_{\alpha}$. After predicting the means of these variables, the denoised versions of the variables at a timestep $T-1$ are obtained and fed back into the model for an additional $N_{T}$ denoising steps. Finally, the BAE is used to decode the encoded optimized wings into coordinate space. In this work we use a total of $N_{T}$ = 1000 denoising time steps. For the architecture, we use a 1D convolutional U-Net, and condition the model by simply concatenating the inputs together after feeding them through a multilayer perceptron feature embedding component. 

Because the model has denoised outputs with vastly different dimensionalities, we also take care to apply appropriately scaled weightings to the loss function. We choose to weight $W_{opt}$, $\alpha$ and $\eta_{y}$ in a 500:1:9 ratio, roughly proportional to their dimensionalities - achieved through a coarse manual tuning. For the variance schedules, we set $\beta_{end}$ = 0.02, and $\beta_{start}$ = 1e-4 for both $\alpha$ and $\eta_{y}$ as is typically done in literature~\cite{ho2020denoising}. However, because we are interested in capturing relatively small changes in wing shape, we set $\beta_{start}$ = 1e-6 for $W_{opt}$, which we found helps improve sample quality, as smaller details are captured more accurately. Finally, our model was trained using 661 of the 776 wings, with 38 wings being used as the validation dataset, and 77 being reserved for the test set. Our model has 1.95M parameters, and was trained for 20000 epochs, a process which took about 20 minutes on an NVIDIA 4070 GPU. 

\subsection{Metrics}

In this work several metrics are used to assess the quality of model generations:

\subsubsection{Maximum Mean Discrepancy}
We make use of maximum mean discrepancy (MMD) to capture the differences between model and data distributions. MMD specifically attempts to quantify the similarity between two distributions, and is defined as:

\begin{equation}
\text{MMD}^2(P, Q) = \mathbb{E}_{x, x' \sim P}[k(x, x')] + \mathbb{E}_{y, y' \sim Q}[k(y, y')] - 2\mathbb{E}_{x \sim P, y \sim Q}[k(x, y)].
\end{equation}
Here, $P$ and $Q$ represent two distributions we are interested in comparing. $x$ and $y$ are independent random variables from the two aforementioned distributions. $x$ and $y$ have copies in the form of $x'$ and $y'$ from the same $P$ and $Q$ distributions. Finally, we choose a kernel function, $k$ to capture the differences between individual dimensions in our distributions. 

\subsubsection{Volume Constraint Satisfaction}
In order to quantify the feasibility of the generated designs, we define a simple metric that measures the percent of generated designs which satisfy the volume constraint, i.e:

\begin{equation}
100 \times \frac{1}{N_{\textit{gen}}}\sum_{i=1}^{N_{\textit{gen}}}n_{i}\begin{cases}
n_{i}=0 & \text{if}\;V_{i,\textit{gen}} < V_{min} \\
n_{i}=1 & \text{if}\;V_{i,\textit{gen}} \geq  V_{min} \\
\end{cases}
\end{equation}

\subsubsection{Mean Squared Error}
We compute mean squared error (MSE) for both shape and angle of attack. MSE is defined as follows:
\begin{equation}
\text{MSE} = \frac{1}{N}\sum_{i=1}^{N}(y_i - \hat{y}_i)^2, 
\end{equation}
where $y_{i}$ is the ground truth value, and $\hat{y_{i}}$ is the predicted or generated value. 

\subsubsection{Vendi Score}

To quantify diversity of the generated designs, we use Vendi Score~\cite{vendi}, which is a measure of the number of unique elements in a set based on their similarity, according to a kernel function. The Vendi Score is~\cite{vendi}:
\begin{equation}
\text{VS}_{k}(X) = \exp\left(-\sum_{i=1}^{n} \lambda_i \log \lambda_i\right) 
\end{equation}
Where $X = \{x_1, x_2, \ldots, x_n\}$ is the set of generated designs, $\mathbf{K}$ is the normalized similarity matrix and $\lambda_i$ are the eigenvalues of $\mathbf{K}$. The Vendi score ranges from 1, meaning all the designs are the same, to $n$, meaning all the designs are unique, with $n$ being the total number of samples being scored. We use Vendi score, primarily because it was a more numerically stable and interpretable metric than other existing diversity metrics (e.g determental point processes) typically employed in literature. 

Note that in our metrics, normalized Vendi score refers to the Vendi score normalized by the ground-truth Vendi score, where the ground truth Vendi score is computed for each dataset split. This makes it more clear how well the model captures the diversity of the original dataset, with values $<1$ being less diverse, and values $>1$ being more diverse. 

\subsubsection{Kernel Parameters}
Both MMD and Vendi score rely heavily on the choice of kernel, and the kernel's parameters. In this work, we choose a Gaussian kernel function~\cite{gretton2012kernel}:

\begin{equation}
k(x, y) = \exp\left(-\gamma \|x - y\|^2\right),
\end{equation}
with bandwidth parameter $\gamma$. 

Since shape and angle of attack are distributed differently, we choose different $\gamma$ for each. We average the statistics over $\gamma = [0.5, 25, 50, 100]$ for both angle of attack and spanwise wing shape. In both cases, the metrics are computed for each $\gamma$ and then averaged.

\subsubsection{Note: Spanwise Metrics}
For most metrics reported, we chose to compute metrics on each individual airfoil section on the 3D wing and then average the result across the sections. This is opposed to directly computing the metrics on the entire 3D wing. We chose to do this because we found that the metrics could become uninterpretable in the relatively high-dimensional wing space. 

\section{Generative Model Results}

Model performance was evaluated on the held-out test set using the metrics defined above. Statistics were computed from 10 forward passes over each dataset split. Table~\ref{tab:EvalMetrics_baseline} reports model performance on the training, validation and test set. The statistics are derived from 10 passes over each dataset split. 

\begin{table}[H]
\centering
\caption{Evaluation metrics across training, validation, and test sets.}
\label{tab:EvalMetrics_baseline}
\resizebox{\columnwidth}{!}{%
\begin{tabular}{llllllll}
\textbf{Dataset} & \textbf{MSE} & \textbf{MMD (Spanwise Avg.)} & \textbf{MSE ($\alpha$)} & \textbf{MMD ($\alpha$)} & \textbf{Vendi (Spanwise Avg.)} & \textbf{Vendi (Normalized)} & \textbf{Vol. Constraint (\%)} \\ \hline
Training   & 1.3E-05 $\pm$ 0.0E+00 & 0.074 $\pm$ 0.012 & 0.071 $\pm$ 0.004 & 0.026 $\pm$ 0.007 & 64.59 $\pm$ 8.33 & 0.550 $\pm$ 0.071 & 99.83 $\pm$ 0.11 \\ \hline
Validation & 1.7E-05 $\pm$ 3.0E-06 & 0.000 $\pm$ 0.000 & 0.102 $\pm$ 0.018 & 0.003 $\pm$ 0.006 & 17.75 $\pm$ 0.64 & 0.856 $\pm$ 0.031 & 98.95 $\pm$ 1.29 \\ \hline
Test       & 3.1E-05 $\pm$ 2.1E-05 & 0.044 $\pm$ 0.063 & 0.203 $\pm$ 0.018 & 0.007 $\pm$ 0.010 & 27.58 $\pm$ 2.58 & 0.803 $\pm$ 0.075 & 99.74 $\pm$ 0.52
\end{tabular}
}%
\end{table}

We interpret the statistics in Table~\ref{tab:EvalMetrics_baseline} as indicating a relatively good agreement with the ground-truth data. In general, the results also indicate that the model appears to have more difficulty in angle of attack generation than in shape reconstruction, given the relatively low shape reconstruction MSE. This may be because angle of attack must represent nonlinear behavior in a single scalar variable, whereas shape has more degrees of freedom to distribute the same error. This could also be misleading in terms of measuring model performance, as smaller changes in shape can result in proportionality larger changes in performance than with angle of attack. In addition, despite the low error, the normalized Vendi score indicates that the model is still measurably less diverse than the ground-truth data. Finally, we note that in certain cases, where the dataset is small, the MMD value can collapse to 0, as is the case for the validation dataset. 

We also examined correlations between input conditions and model errors on the test set ($N=77$). No statistically significant correlations were observed at the 0.05 level, although weak direct trends were noted between shape MSE and minimum volume constraint ($\rho=0.21$) and between angle of attack MSE and Reynolds number ($\rho=0.21$). A larger test set would be needed to determine whether these trends are robust.

\begin{figure}[H]\centering
\includeinkscape[width=0.85\textwidth]{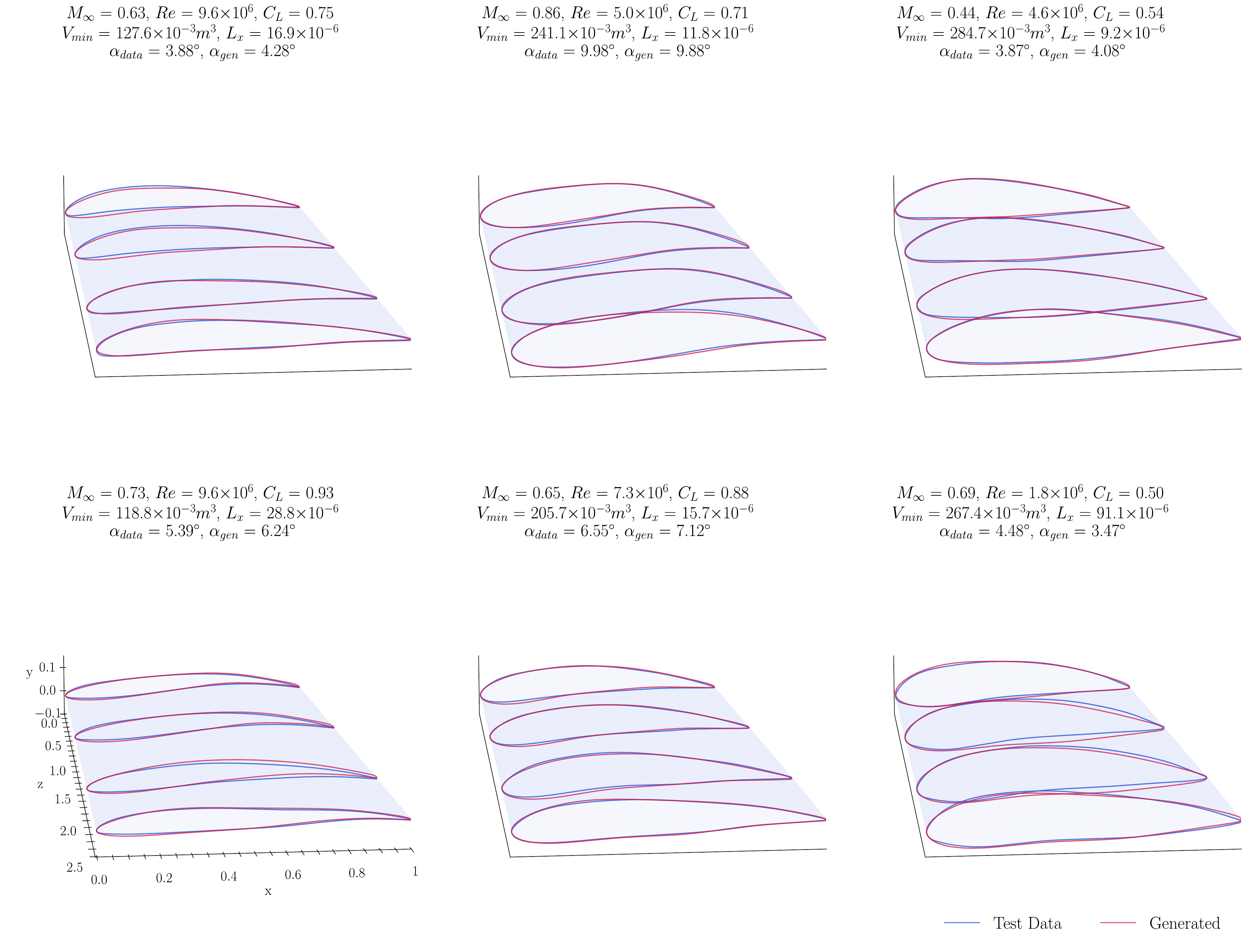}
\caption{Generated and Ground Truth Optimized Wings\protect\footnotemark}
\label{fig:test_gt_gen}
\end{figure}

Qualitatively, we present sample model generations in Figure~\ref{fig:test_gt_gen}. Note that in the Figure we have also included, $L_{x}$, the wing shape MSE between the ground-truth and model generated sample, as a point of reference.

\begin{figure}[H]
    \centering
    \begin{subfigure}[b]{0.49\textwidth}
        \centering
        \includeinkscape[width=\textwidth]{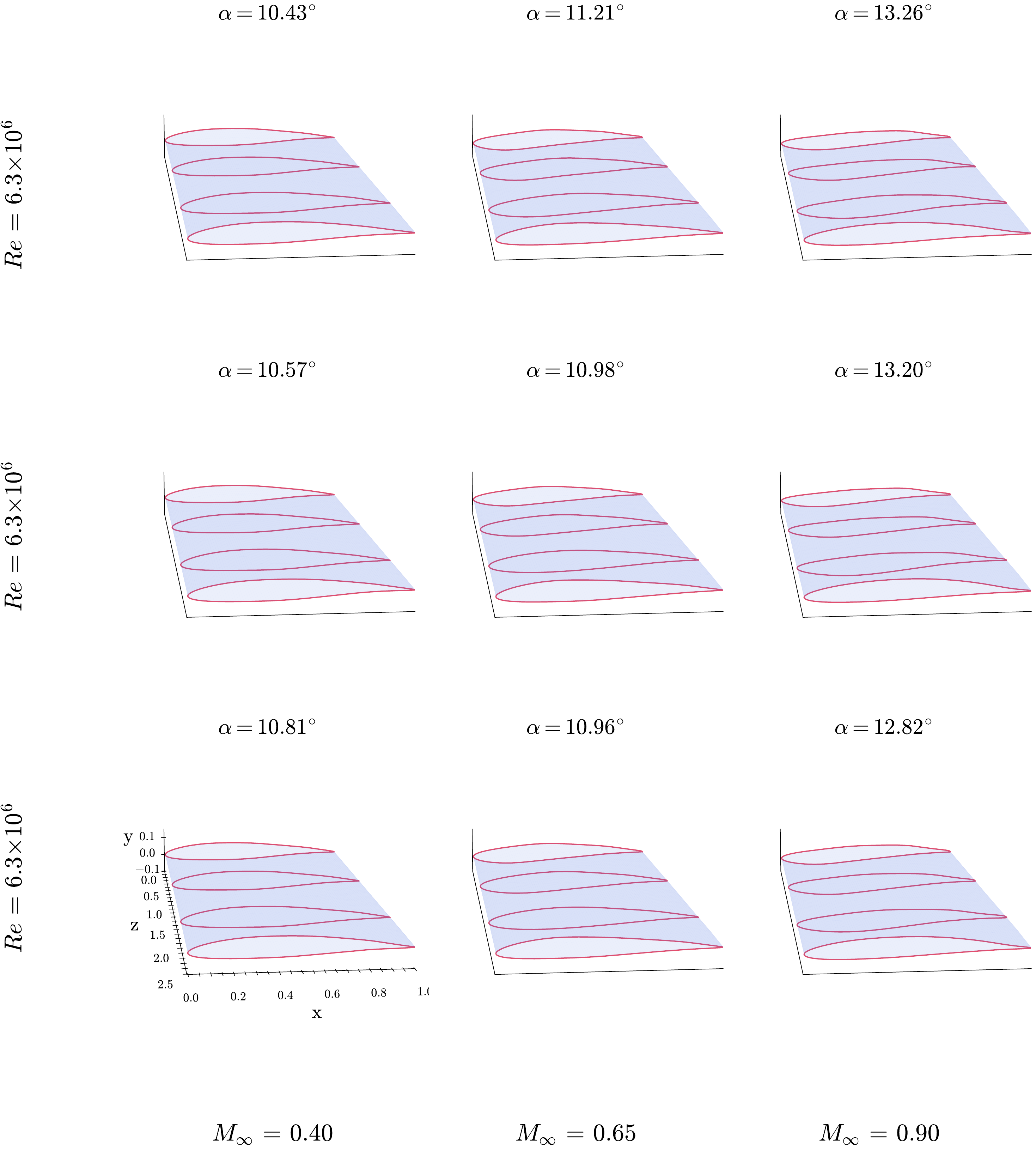}
        \caption{Mach and $C_{L}$ Variation}
        \label{fig:gen_mach_re_grid_subfig}
    \end{subfigure}
    \hfill
    \begin{subfigure}[b]{0.49\textwidth}
        \centering
        \includeinkscape[width=\textwidth]{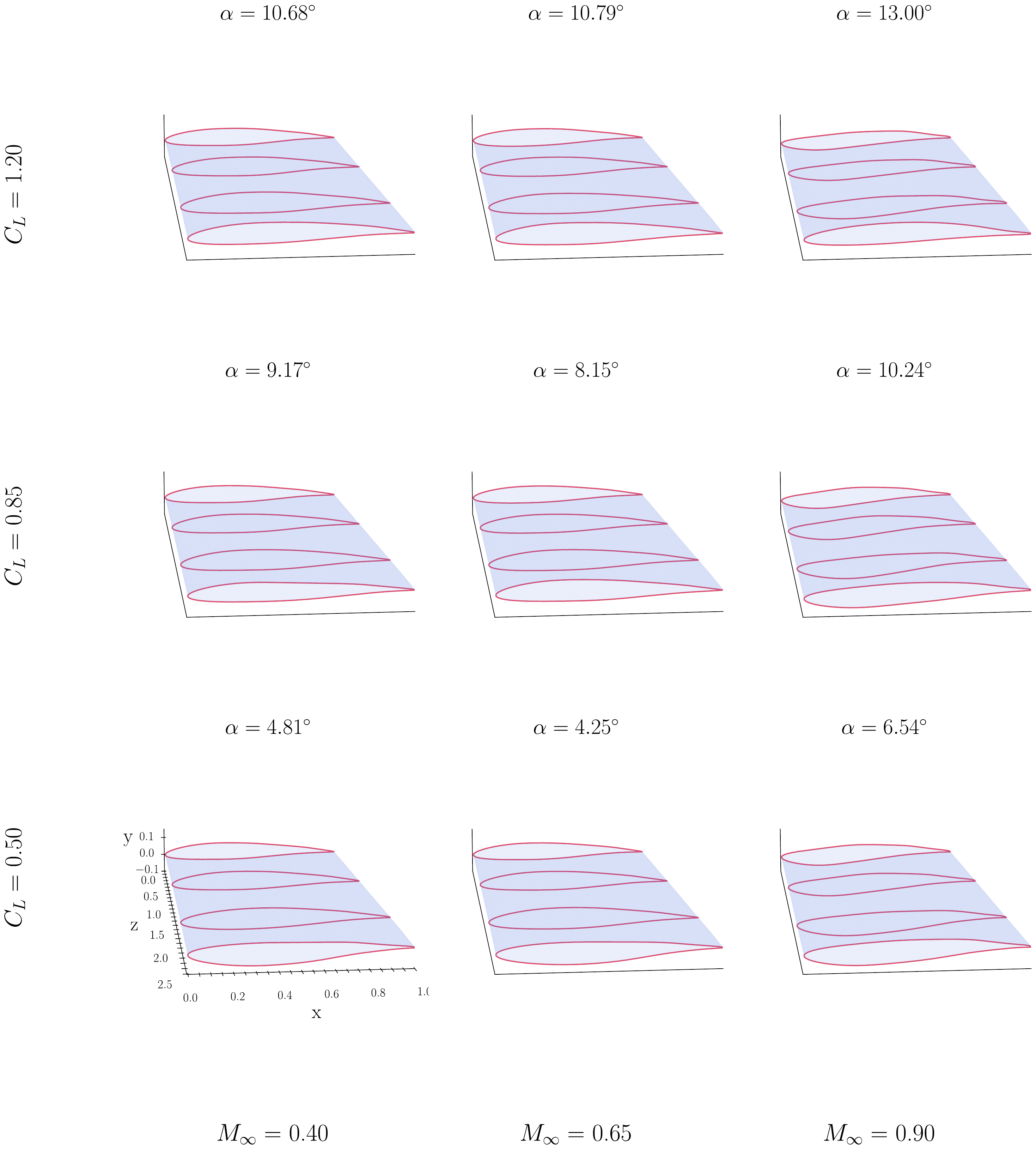}
        \caption{Mach and Reynolds Variation}
        \label{fig:gen_mach_CL_grid_subfig}
    \end{subfigure}
    \vfill
    \begin{subfigure}[b]{0.16\textwidth}
        \centering
        \includeinkscape[width=\textwidth]{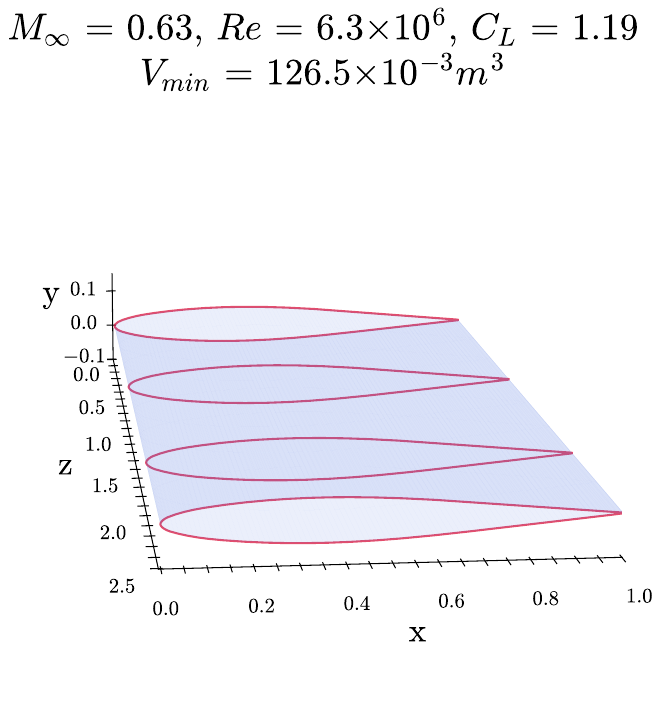}
        \caption{Initial Design}
        \label{fig:gen_mach_CL_grid_init_wing}
    \end{subfigure}
        \caption{Generated Designs for Different Parameters}
        \label{fig:gen_grids}
\end{figure}

We now visually examine how model output varies with lift coefficient, Reynolds number and Mach number, in Figure~\ref{fig:gen_grids}. 

We observe that higher Mach numbers seem to result in more  supercritical-type airfoil cross sections. Qualitatively, Mach number appears to have the greatest overall impact on wing cross-sectional shape, followed by the coefficient of lift, and finally Reynolds number, which has a noticeably decreased impact on shape compared to the other two parameters. 

\begin{figure}[H]
    \centering
    \begin{subfigure}[b]{0.35\textwidth}
        \centering
        \includeinkscape[width=\textwidth]{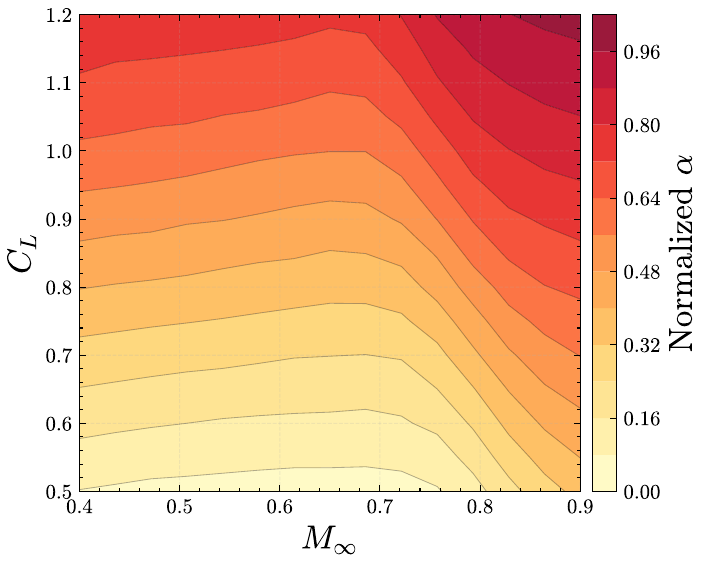}
        \caption{Mach and $C_{L}$: Relative Mean}
        \label{fig:aoa_mach_cl_mean_subfig}
    \end{subfigure}
    \begin{subfigure}[b]{0.35\textwidth}
        \centering
        \includeinkscape[width=\textwidth]{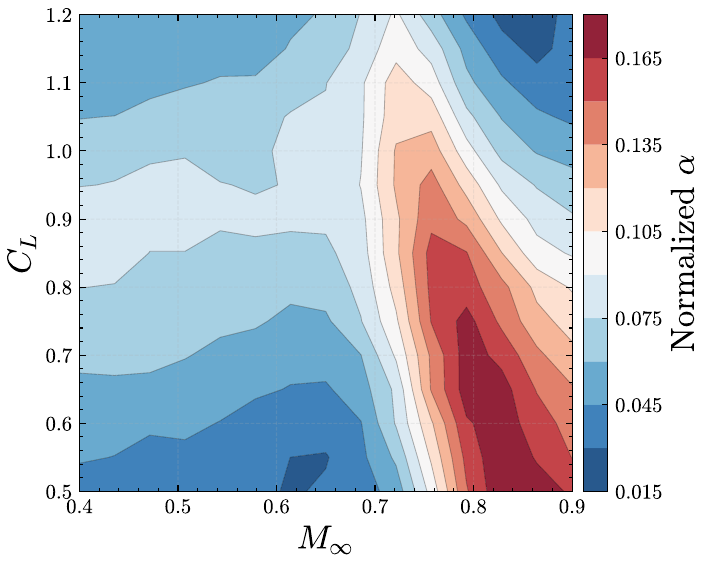}
        \caption{Mach and $C_{L}$: Relative Deviation}
        \label{fig:aoa_mach_cl_std_subfig}
    \end{subfigure}
    \vfill

    \begin{subfigure}[b]{0.35\textwidth}
        \centering
        \includeinkscape[width=\textwidth]{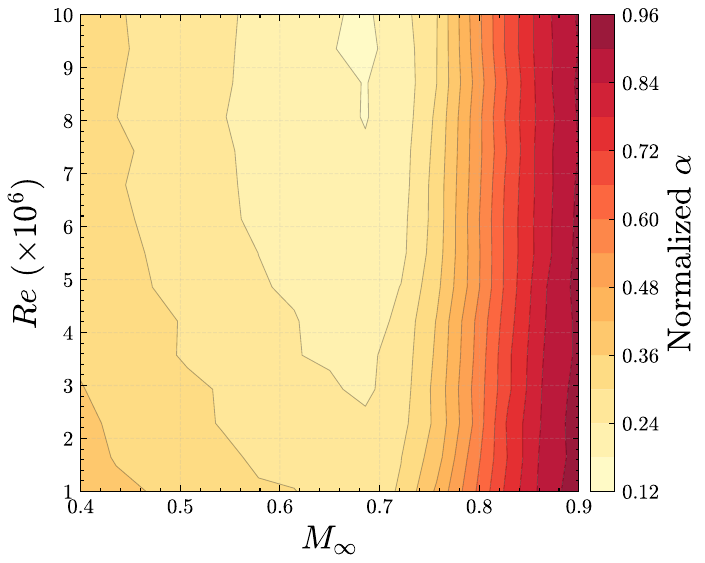}
        \caption{Mach and Reynolds: Mean}
        \label{fig:aoa_mach_re_mean_subfig}
    \end{subfigure}
    \begin{subfigure}[b]{0.35\textwidth}
        \centering
        \includeinkscape[width=\textwidth]{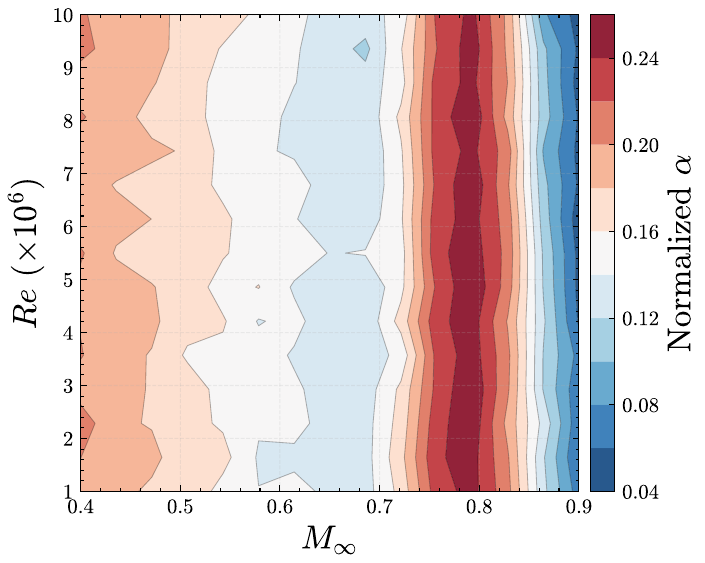}
        \caption{Mach and Reynolds: Relative Deviation}
        \label{fig:aoa_mach_re_std_subfig}
    \end{subfigure}
        \caption{Angle of Attack Variation}
        \label{fig:aoa_variation}
\end{figure}

To better understand how flow parameters effect the predicted angle of attack, we constructed aggregate contour plots across all samples in the dataset. For each sample, the angle of attack was normalized to its minimum and maximum values. Figure~\ref{fig:aoa_variation} displays the mean and standard deviation of these normalized values.

The Mach-$C_{L}$ contours (Figs.~\ref{fig:aoa_mach_cl_mean_subfig}-\ref{fig:aoa_mach_cl_std_subfig}) show that higher $C_{L}$ corresponds to higher $\alpha$ across all Mach numbers, as might be expected. The Mach number dependence is non-monotonic; that is the $\alpha$ required for a given $C_{L}$ constraint decreases as Mach number increases from approximately 0.4 to 0.7, with the trend then reversing at higher Mach numbers, as compressibility effects become more significant (presuming accuracy in the generated results). Variance is highest in the transonic regime (where Mach~$\sim$0.75-0.9). 

The Mach-Reynolds contours (Figs.~\ref{fig:aoa_mach_re_mean_subfig}-\ref{fig:aoa_mach_re_std_subfig}) indicate that beyond a Mach$~\sim$0.8, the predicted $\alpha$ increases with Mach. The minimum average $\alpha$ occurs near a Reynolds number of 1E7 and a Mach number of 0.7. Variance is again concentrated near a Mach number of around 0.8. These nonlinear relationships in the transonic regime, help explain why $\alpha$ may be a relatively difficult learning task. 

\begin{figure}[H]\centering
\includeinkscape[width=0.70\textwidth]{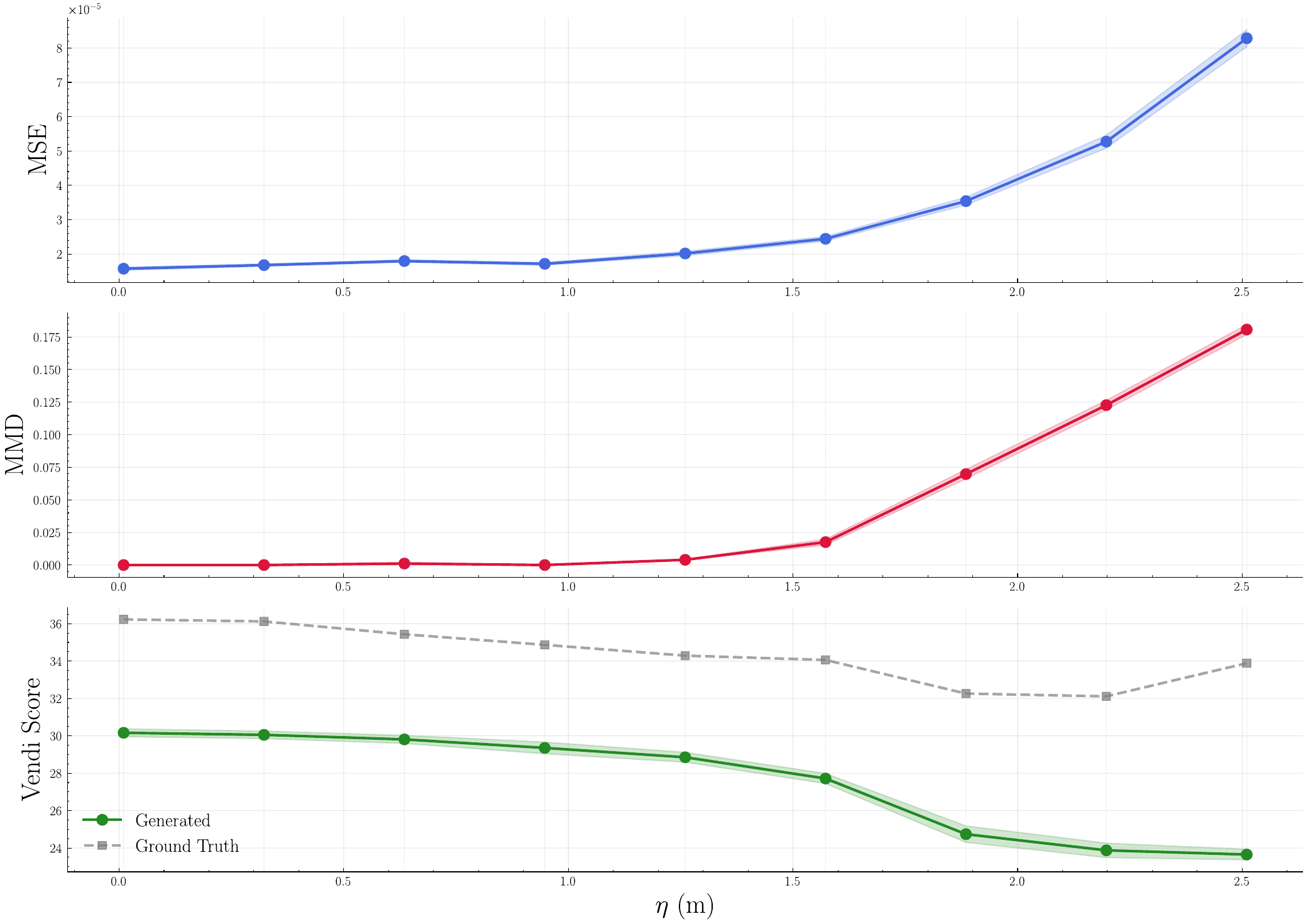}
\caption{Spanwise Metrics}
\label{fig:spanwise_metrics}
\end{figure}

We also briefly explore the relationship between spanwise position and model metrics in Figure~\ref{fig:spanwise_metrics}.
         
It is clear from Figure~\ref{fig:spanwise_metrics} that all metrics are disproportionately impacted towards the last quarter of the span. We observe that MSE and MMD both increase, indicating greater difficulty in capturing wingtip adjacent features. This aligns with our finding that optimization-driven shape changes are largest near the tip, suggesting the model struggles most where the design space is most complex. We also observe diversity, as measured by Vendi score, decreasing towards the tip, although this follows the same trend as in the ground-truth data.

\subsection{Data Ablation Study}

\begin{figure}[H]
    \centering
    \begin{subfigure}[b]{0.495\textwidth}
        \centering
        \includeinkscape[width=\textwidth]{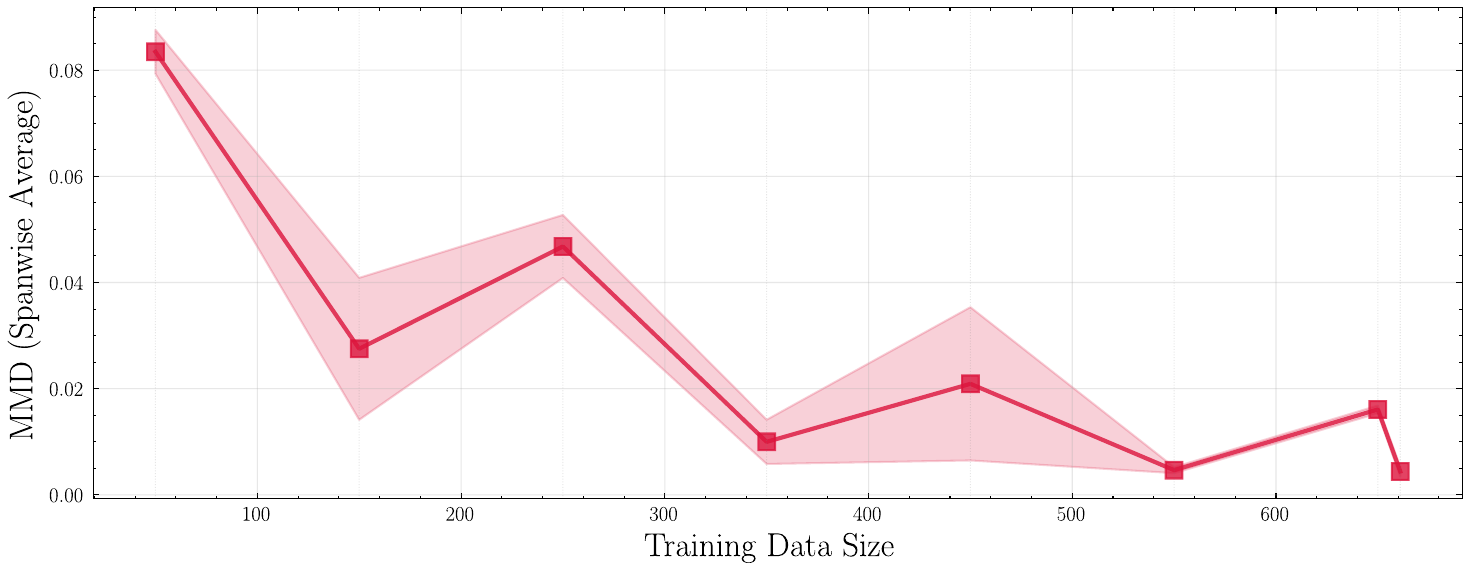}
        \caption{MMD (Wing Shape)}
        \label{fig:ablation_mse}
    \end{subfigure}
    \hfill
    \begin{subfigure}[b]{0.495\textwidth}
        \centering
        \includeinkscape[width=\textwidth]{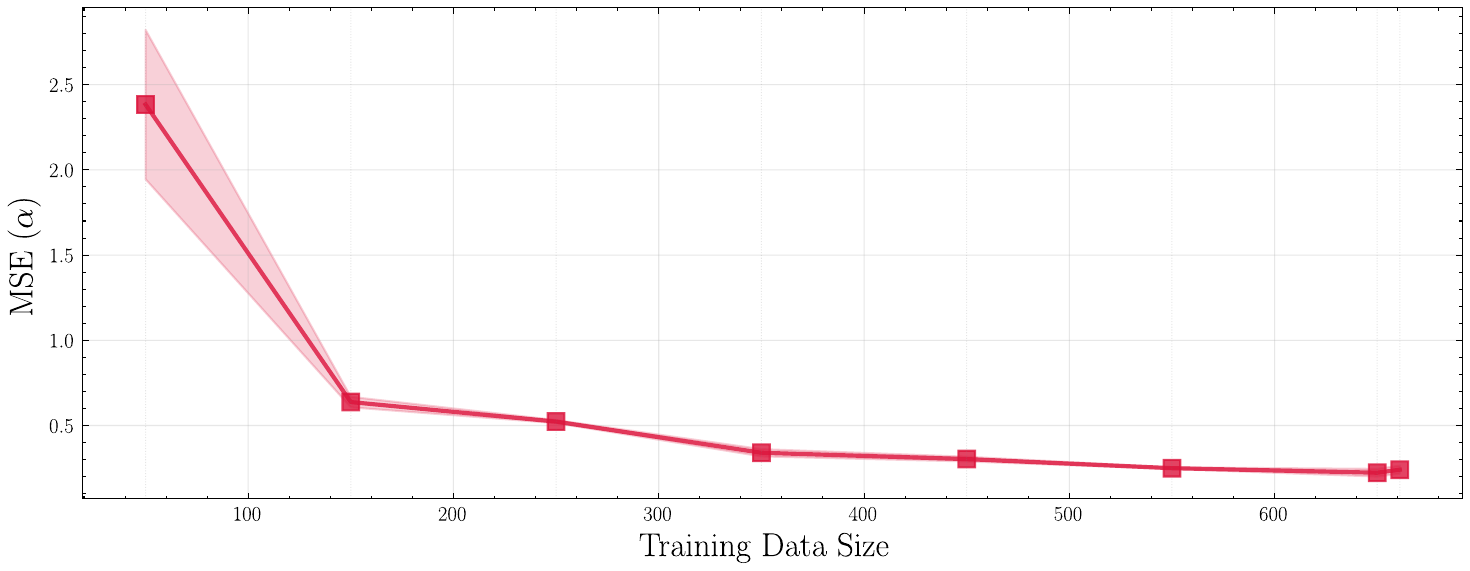}
        \caption{MSE (Angle of Attack)}
        \label{fig:ablation_aoa_mse}
    \end{subfigure}
    \hfill
    \begin{subfigure}[b]{0.495\textwidth}
        \centering
        \includeinkscape[width=\textwidth]{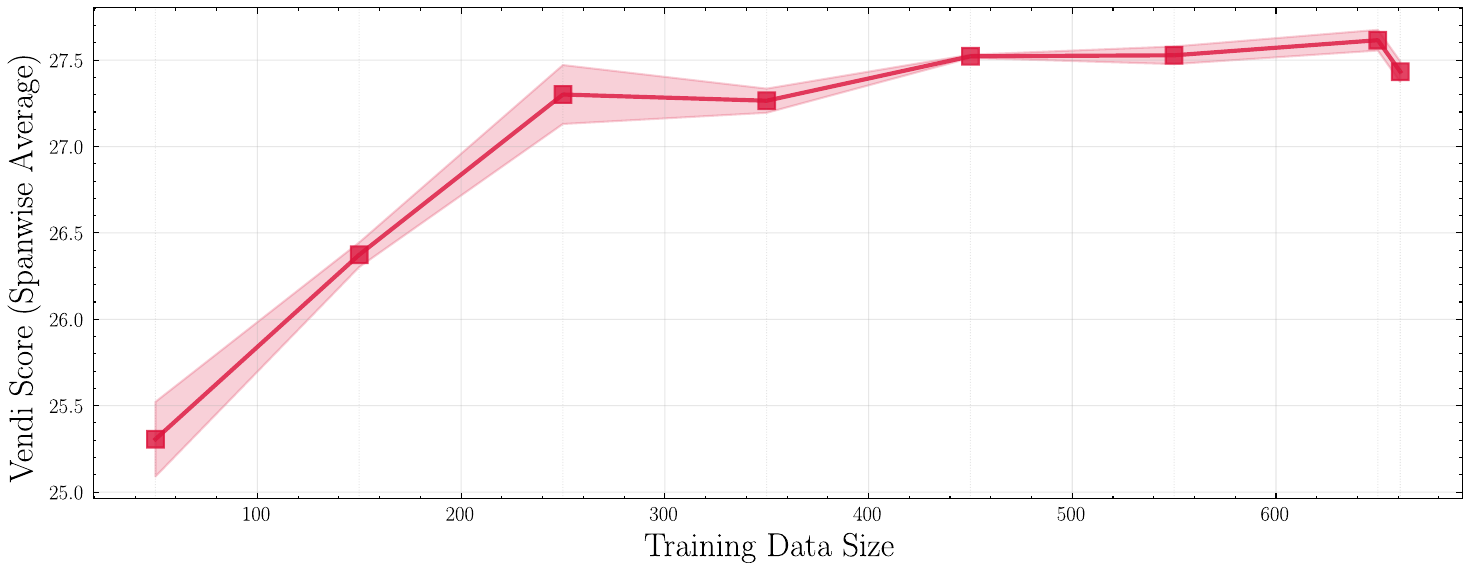}
        \caption{Vendi Score}
        \label{fig:ablation_vendi}
    \end{subfigure}
    \hfill
    \begin{subfigure}[b]{0.495\textwidth}
        \centering
        \includeinkscape[width=\textwidth]{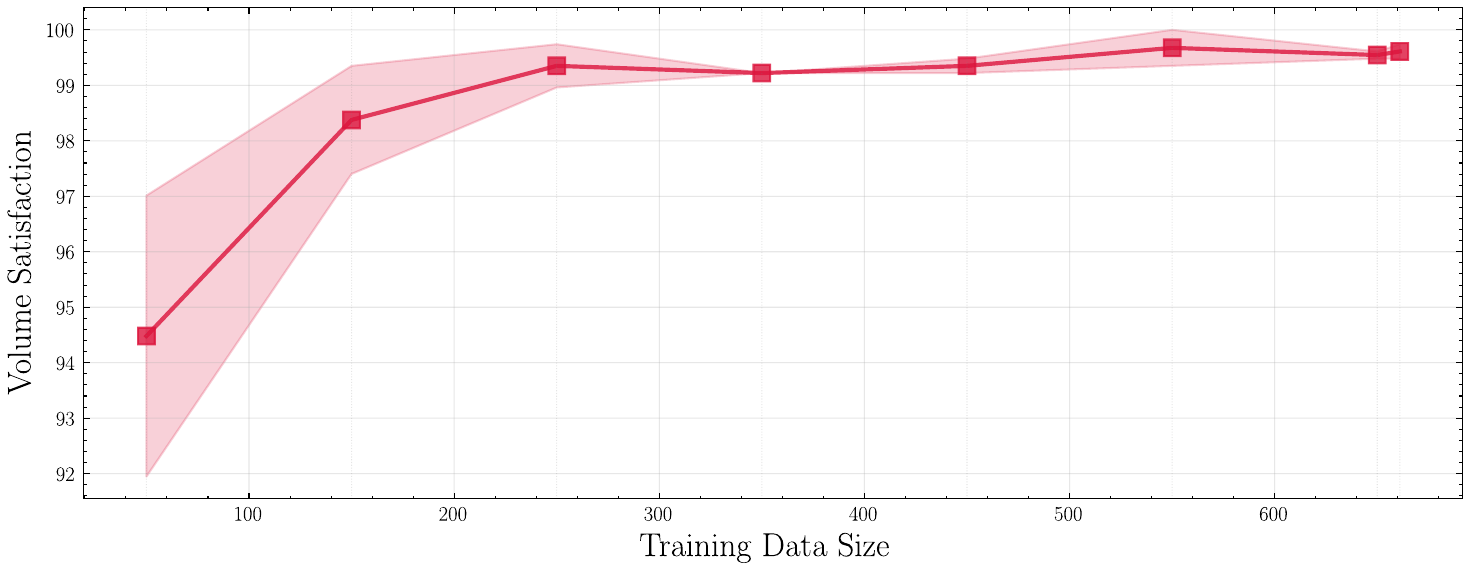}
        \caption{Volume Satisfaction}
        \label{fig:ablation_volume}
    \end{subfigure}
        \caption{Ablation Study: Selected Metrics (Test Data)}
        \label{fig:ablation_study}
\end{figure}

We conduct a data ablation study to determine how model performance scales with training dataset size. We plot four representative metrics as a function of increasing training data in Figure~\ref{fig:ablation_study}.

In the study, each metric evaluation was the result of 10 passes over the test dataset. In addition, 2 models were trained for each dataset size, and their results were averaged for a more accurate comparison. 

To quantify diminishing returns, we computed the relative marginal improvement rate (percent change per additional training sample) for each metric. In this case, the rate is relative the value of the metric computed for the lowest number of training samples, $N=50$. For angle-of-attack MSE, relative improvement rates begin at approximately 0.7\% per sample between 50 and 100 samples, and decline to less than 0.1\% per sample beyond 250 samples, and fall below 0.05\% per sample—or become negative—past 450 samples. Volume constraint satisfaction and normalized Vendi score both show marginal relative improvement rates below 0.01\% per sample beyond 350 samples. Wing shape MMD exhibits high variance throughout, rendering improvement rate estimates inconclusive for this metric. Collectively, these results suggest that approximately 350 training samples are sufficient to capture most learnable structure in the dataset, with additional data yielding negligible performance gains. Despite this, wing shape MMD may still improve measurably beyond this point. We also highlight that the point of diminishing returns for the model implemented in this work may be significantly different for other models, due to implementation related factors. For example, the reconstruction error on the latent control point space may prove to be a bottleneck for metric improvement.

\section{Conclusion and Future Work}

We have presented OptiWing3D, the first publicly available dataset of gradient-based optimized 3D wing designs, addressing a gap in benchmarks for data-driven aerodynamic design. The dataset consists of 1552 simulations consisting of 776 optimized wing designs (with initial and optimized simulations totaling 1552 cases) across realistic subsonic and transonic flight conditions (M = 0.4–0.9, Re = 1E6-1E7). Additionally, OptiWing3D is paired to a dataset of 2D optimized airfoil designs, allowing for direct comparisons between 2D and 3D optimizations and simulations. The dataset was collected using a high-fidelity, structured, RANS-based solver applied to a diverse set of wing geometries. Analysis of the dataset reveals that 3D optimizations diverge most significantly from their 2D counterparts near the wingtip, where three-dimensional aerodynamic effects are strongest. Finally, in order to provide a baseline for future research, we have implemented a conditional latent diffusion model capable of generating optimized wing geometries from flow conditions and optimizer constraints, and we demonstrate that this model achieves shape reconstruction MSE on the order of 1E-5 and captures approximately 80\% of the ground-truth design diversity. A data ablation study reveals that approximately 350 training samples are sufficient to achieve near-optimal performance under the current diffusion model architecture, providing coarse guidance for future efforts.

Several extensions would enhance the dataset's utility. Incorporating global design variables (e.g twist, dihedral, taper) and structural constraints (such as bending moment limits) would enable more realistic design studies. The current dataset is also weighted toward lower Mach numbers due to solver convergence issues; future versions could employ adaptive numerical strategies to improve coverage at higher Mach numbers. Additionally, the assumption of constant freestream temperature becomes less realistic at higher Mach numbers, where adiabatic heating effects are more significant; an improved dataset could feature temperature as an additional sampled parameter. The paired 2D-3D structure also lays groundwork for multi-fidelity transfer learning studies, which we leave for future work.

The OptiWing3D dataset, including wing geometry slices, aerodynamic coefficients, and surface pressure distributions is made available at \url{https://github.com/cashend/OptiWing3D} to support continued research into data-driven aerodynamic shape design.

\section*{Acknowledgments}

\bibliography{sample}

\section{Appendix}

\begin{appendices}

\section{Off-Wall Distance Estimation}

The first cell height for boundary layer is coarsely estimated to achieve a target $y^+$ value using the following variables:

\begin{itemize}
    \item $M$ -- Freestream Mach number
    \item $\text{Re}$ -- Reynolds number
    \item $T_\infty$ -- Freestream temperature (K)
    \item $L_{\text{ref}}$ -- Reference length (m)
    \item $y^+$ -- Target dimensionless wall distance
    \item $R$ -- Specific gas constant for air (287 J/(kg·K))
    \item $\gamma$ -- Ratio of specific heats (1.4 for air)
    \item $\mu_0$ -- Reference viscosity (1.716 $\times$ 10$^{-5}$ Pa·s)
    \item $T_0$ -- Reference temperature for Sutherland's law (273.15 K)
    \item $S$ -- Sutherland's constant (110.4 K)
    \item $a$ -- Speed of sound (m/s)
    \item $u$ -- Freestream velocity magnitude (m/s)
    \item $\mu$ -- Dynamic viscosity (Pa·s) derived from Sutherland's law
    \item $\rho$ -- Freestream density (kg/m$^3$)
    \item $C_f$ -- Skin friction coefficient (empirical correlation)
    \item $\tau_w$ -- Wall shear stress (Pa)
    \item $u_\tau$ -- Friction velocity (m/s)
    \item $\delta$ -- Off-wall distance for first cell (m)
\end{itemize}

With Mach, Reynolds number and temperature known, the other variables are computed using Sutherland's law. The following equation is then used to estimate the $y+$ value:

\begin{equation}
\delta = \frac{y^+ \mu}{\rho \sqrt{\tau_w/\rho}} = \frac{y^+ \mu}{\sqrt{\rho \tau_w}}
\end{equation}

where $\tau_w = C_f \cdot \frac{1}{2} \rho u^2$, with $u = M\sqrt{\gamma R T_\infty}$ and $\mu$ computed via Sutherland's law. $C_f$ is approximated with the Prandtl-Schlichting estimate (valid between $10^5 < \text{Re} < 10^9$) for turbulent flat-plate boundary layers~\cite{schlichting2016boundary}:
\begin{equation}
C_f = \left(2\log_{10}(\text{Re}) - 0.65\right)^{-2.3}
\end{equation}

\end{appendices}

\end{document}